\documentclass[11pt,a4paper]{article}

\usepackage[utf8]{inputenc}
\usepackage[T1]{fontenc}
\usepackage{geometry}
\usepackage{amsmath,amssymb,amsfonts}
\usepackage{graphicx}
\usepackage{booktabs}
\usepackage{url}
\usepackage[colorlinks=true, linkcolor=blue, citecolor=blue, urlcolor=blue]{hyperref}
\usepackage{anyfontsize}
\usepackage{colortbl}   
\usepackage{xcolor}     
\usepackage{steinmetz}
\usepackage{setspace} 
\usepackage{amsmath,amssymb,amsthm,mathtools} 
\usepackage{bm} \usepackage{graphicx} 
\usepackage{booktabs} 
\usepackage{multirow} 
\usepackage{enumitem} 
\usepackage{hyperref} 
\usepackage{varwidth}
\usepackage{bbold} 
\usepackage{chngcntr} 

\newcommand{\Bias}{\operatorname{B}}  
\newcommand{\Biask}{\widehat{\Bias}_k}
\newcommand{\Biast}{\widehat{\Bias}_{\mathrm{total}}}
 
\newcommand{\Var}{\operatorname{Var}} 
\newcommand{\Prob}{\operatorname{P}} 

\theoremstyle{definition} 
\newtheorem{definition}{Definition} 
\theoremstyle{plain} 
\newtheorem{lemma}{Lemma}
\newtheorem{proposition}{Proposition}
\newtheorem{theorem}{Theorem}
\newtheorem{corollary}{Corollary}
\theoremstyle{remark} 
\newtheorem{remark}{Remark}
\newcommand{\bmsection}{\section}
\newcommand{\bmsubsection}{\subsection}
\newcommand{\bmsubsubsection}{\subsubsection}
\title{\textbf{Measuring Kaplan--Meier Overestimation Under Competing Risks: Event Ordering, Quantification, and Exact Inference}}

\author{
  \textbf{Ferran Espuny-Pujol} \\
  Department of Computer Science \\
  University of Reading, UK \\
  \texttt{f.espuny-pujol@reading.ac.uk}
}
\date{\today}

\begin{document}
\maketitle

\begin{abstract}
	Competing risks arise when individuals may experience multiple mutually exclusive event types. 
	Competing events are frequently treated as censoring and analysed 
	using Kaplan--Meier (KM) methods, which overestimate real-world event risks. 
	While the existence of KM overestimation is well known, 
	its determinants, magnitude, and inferential properties remain poorly characterised.
	
	We develop a nonparametric framework that 
	treats competing risks overestimation as a statistical quantity in its own right. 
	We show that competing event frequency alone is insufficient to characterise KM overestimation 
	and develop an event ordering theory describing how temporal event allocation influences overestimation. 
	Based on these results, we derive a closed-form Total Overestimation Diagnostic (TOD) 
	that quantifies total KM overestimation directly from routine survival analysis output. 
	We establish exact error equivalence results for both cause-specific and total overestimation. 
	In particular, Product Limit Error Equivalence shows that inference for TOD 
	reduces exactly to ordinary overall survival estimation.
	
	Simulation experiments validate the theoretical results and 
	show that independent censoring primarily affects precision rather than overestimation magnitude. 
	Across six benchmark competing risks datasets, cause-specific overestimation 
	is characterised by both total overestimation and temporal event allocation. 
	These results provide a practical framework for assessing competing risks relevance and 
	determining when cumulative incidence methods should be preferred to na\"ive KM risks.

	The proposed framework transforms competing risks overestimation 
	from a qualitative warning into a measurable quantity that can be 
	routinely quantified, interpreted, and incorporated into reporting practice.
\end{abstract}

\section{Introduction}

	Competing risks occur when individuals may experience multiple mutually exclusive event types, 
	such that the occurrence of one event prevents the occurrence of the others 
	\cite{prentice1978analysis, kalbfleisch2002statistical, geskus2016data}.
	Examples occur frequently in clinical and epidemiological research, including 
	relapse versus non-relapse mortality following haematopoietic stem cell transplantation 
	\cite{gooley1999estimation}, 
	progression to plasma cell myeloma versus death without prior progression in studies of 
	monoclonal gammopathy of undetermined significance (MGUS) 
	\cite{kyle1993benign, kyle2002review}, 
	and numerous applications in oncology, cardiology, and ageing research 
	\cite{austin2016introduction, geskus2016data, koller2011competing}.
	In these settings, investigators are often interested in probabilities of clinically observable events, such as
	relapse within five years, 
	disease progression before death, or
	death from a specific cause. 
	These quantities are naturally described by
	cumulative incidence functions \cite{pepe1993kaplan}.
	Their corresponding nonparametric estimator is the Aalen--Johansen (AJ) estimator 
	\cite{aalen1978empirical, johansen1978empirical}.
	
	Competing risks are frequently ignored in routine practice.	 
	Reviews document frequent use of Kaplan--Meier (KM) methods 
	in settings where cumulative incidence is the quantity of interest 
	\cite{austin2017accounting,koller2011competing}.
	The usual approach treats competing events as censoring and 
	estimates event probabilities via the complement of the KM estimator. 
	When competing events are treated as censoring, 
	$1-\hat{S}_{\mathrm{KM},k}(t)$ does not estimate the cumulative incidence of event type $k$. 
	Instead, it estimates a net probability corresponding to a hypothetical setting 
	in which competing events do not occur \cite{pepe1993kaplan,pintilie2006competing}. 
	As a result, KM estimation generally overstates event probabilities.
	For example, when estimating relapse following transplantation, 
	deaths without relapse preclude subsequent relapse.
	Treating such deaths as censoring therefore 
	evaluates the relapse risk that would have been observed 
	had those deaths not occurred \cite{gooley1999estimation}. 
	
	Although KM overestimation under competing risks is well recognised, 
	its underlying structure remains poorly understood. 
	The methodological literature provides extensive guidance 
	on the analysis and interpretation of competing risks 
	\cite{pepe1993kaplan, gooley1999estimation, andersen2012competing, austin2017accounting}. 
	Relatively little attention has been devoted to understanding 
	the magnitude, determinants, and practical consequences of KM overestimation. 
	In particular, the role of event ordering remains poorly understood. 
	Direct comparison of KM and AJ estimates can reveal the presence of overestimation, 
	but does not explain why overestimation occurs, 
	what governs its magnitude, 
	how it accumulates over time, or 
	how uncertainty should be quantified. 
	Competing event frequency is an obvious determinant of overestimation. 
	However, datasets with similar competing event frequencies may exhibit 
	markedly different levels of KM overestimation.
	This suggests that competing event frequency alone 
	is insufficient for assessing competing risks relevance 
	and raises more general questions regarding the mechanisms 
	and practical consequences of KM overestimation.
	
	The central idea of this paper is
	that KM overestimation under competing risks 
	admits a formal theoretical characterisation through 
	event ordering, quantification, and exact inference.
	We further show that this characterisation provides 
	a principled basis for assessing competing risks relevance.
	Rather than viewing KM overestimation as a
	well recognised consequence of competing risks,
	we treat it as a measurable quantity that can be 
	quantified, interpreted, subjected to formal inference,
	and incorporated into routine reporting. 
	By characterising both the magnitude of overestimation and the mechanisms by which it arises, 
	we develop a unified framework for understanding KM overestimation 
	and assessing the practical relevance of competing risks.
	
	The proposed framework has four complementary components. 
	First, we develop an event ordering theory that characterises 
	how temporal event allocation influences KM overestimation, 
	including attainable bounds, local accumulation mechanisms, 
	and extremal ordering configurations. 
	Second, we introduce two complementary measures related to KM overestimation. 
	The Total Overestimation Diagnostic (TOD) quantifies total KM overestimation, 
	whereas the Order Index (OI) characterises temporal event allocation, 
	which we show is a fundamental determinant of cause-specific KM overestimation. 
	Third, we derive exact inferential results for both cause-specific and total
	overestimation, culminating in the Product Limit Error Equivalence theorem, 
	which links inference for KM overestimation directly to standard KM survival inference.
	
	Finally, we demonstrate how the proposed theory, measures, 
	and inferential results can be used to assess competing risks relevance. 
	Our results show that competing event frequency alone is insufficient. 
	Rather, competing risks relevance depends on both 
	the magnitude of total KM overestimation and 
	the temporal allocation of event types. 
		
	The remainder of the paper is organised as follows.
	Section~\ref{sec:background} introduces the competing risks framework and KM overestimation. 
	Sections~\ref{sec:event_ordering}--\ref{sec:global_diagnostic} develop
		the event ordering theory, cause-specific inference, TOD, and Product Limit Error Equivalence.
	Section~\ref{sec:empirical_evaluation} presents the empirical evaluation and 
	Section~\ref{sec:practical_relevance} develops a practical interpretation framework 
	to assess competing risks relevance in applied settings.
	Section~\ref{sec:discussion} concludes with a discussion of implications, limitations, 
	reporting recommendations, and future directions.
	
\section{Competing Risks and KM Overestimation} 
	\label{sec:background}
	
\subsection{Competing Risks Framework}
	\label{subsec:background}
	
	Consider a competing risks setting with $K\ge2$ mutually exclusive event types. 
	Let $T$ denote the time to the first event and $J\in\{1,\ldots,K\}$ its event type. 
	Let $C$ denote the right censoring time. 
	The observed data are $X=\min(T,C)$ and $E=J$ if $T\le C$, with $E=0$ otherwise 
	\cite{kalbfleisch2002statistical,geskus2016data,pintilie2006competing}.
	The cause-specific cumulative incidence function plays a fundamental role in competing risks analysis 
	and represents the real-world probability that an event of type $k$ has occurred by time $t$ in the presence of competing events.
	Formally,
	\begin{equation} 
		F_k(t) = \Prob(T \le t, J = k) = \int_0^t S(u^-)\lambda_k(u)\,du, \qquad k=1,\ldots,K, 
		\label{eq:cif_def} 
	\end{equation} 
	where $S(u^-)=\Prob(T\ge u)$ denotes the overall event free survival function and 
	$\lambda_k(u)$ denotes the cause-specific hazard function for event type $k$ 
	\cite{aalen1978empirical,kalbfleisch2002statistical,geskus2016data}. 
	
	A key property of competing risks analysis is the overall \textit{risk decomposition} identity,
	which expresses the probability of experiencing any event type by time $t$
	as the sum of the cause-specific cumulative incidences 
	\cite{kalbfleisch2002statistical,pintilie2006competing,geskus2016data}: 
	\begin{equation} 
		1-S(t) = \sum_{k=1}^{K} F_k(t). 
		\label{eq:risk_decomposition} 
	\end{equation} 
	
	Let $t_1<t_2<\cdots<t_m$ denote the ordered distinct event times across all causes. 
	At each event time $t_i$, 
	$r(t_i^-)$ denotes the number of individuals remaining alive, event free, and under observation (risk set) immediately before $t_i$,
	and $d_k(t_i)$ denotes the number of events of type $k$ occurring at $t_i$, so that  
	$d(t_i)=\sum_{k=1}^{K} d_k(t_i)$ is the total number of events at $t_i$. 
	
	The standard nonparametric estimator of the cumulative incidence function 
	is the {\it Aalen--Johansen (AJ) estimator}\footnote{
		Note that if an event occurs at $t_i$, then necessarily $d(t_i)\ge1$ and therefore
		$r(t_i^-)\ge d(t_i)\ge1$.
		Hence, immediately before every observed event time the risk set is non-empty:
		$r(t_i^-)>0$, $i=1,\ldots,m$. 
	}, 
	\begin{equation} 
		\widehat F_k^{AJ}(t) = \sum_{t_i\le t} 
			\widehat S_{PL}(t_{i-1}) 
			\frac{d_k(t_i)}{r(t_i^-)}, 
		\label{eq:aj_estimator} 
	\end{equation} 
	where 
	\begin{equation} 
		\widehat S_{PL}(t) = \prod_{t_i\le t} 
			\left( 1-\frac{d(t_i)}{r(t_i^-)} \right) 
		\label{eq:pl_estimator} 
	\end{equation} 
	is the overall product limit estimator of event free survival 
	\cite{aalen1978empirical,johansen1978empirical,geskus2016data}. 
	Equation~\eqref{eq:aj_estimator} shows that cumulative incidence accumulates only through events of type $k$, 
	with each increment weighted by the overall event free survival probability immediately beforehand. 
	Competing events therefore influence future cumulative incidence indirectly through 
	their impact on event free survival \cite{kalbfleisch2002statistical,pintilie2006competing, geskus2016data}.
	Inference for cumulative incidence functions is commonly based on 
	counting process methods and large sample theory
	\cite{aalen1978empirical,kalbfleisch2002statistical,geskus2016data}.
	
	In contrast, when competing events are treated as ordinary censoring, 
	the cause-specific {\it Kaplan--Meier (KM) estimator} is \cite{kaplan1958nonparametric}: 
	\begin{equation} 
		\widehat S_{KM,k}(t) = \prod_{t_i\le t} \left( 1-\frac{d_k(t_i)}{r(t_i^-)} \right). 
		\label{eq:km_cause_specific} 
	\end{equation} 
	Its complement, $1-\widehat S_{KM,k}(t)$, is frequently interpreted and reported as an estimate of the probability of event type $k$ 
	\cite{austin2017accounting,koller2011competing}.
	
	\paragraph*{Assumptions.}
	Unless stated otherwise, the results developed throughout this article assume
	a standard competing risks setting with independent right censoring	
	\cite{kalbfleisch2002statistical, andersen1993counting, geskus2016data}. 
	Left truncation (delayed entry), interval censoring, and recurrent event processes are not considered. 

\subsection{Quantifying KM Overestimation} \label{sec:empirical_bias} 
	
	We begin by defining measures of KM overestimation for individual event types and for the competing risks setting as a whole.
	
	\begin{definition}[Empirical Cause-Specific Overestimation]
		For event type $k$, we define the {\it empirical cause-specific KM overestimation} at time $t$ as 
		\begin{equation} 
			 \Biask(t) = \left( 1-\widehat S_{\mathrm{KM},k}(t) \right) - \widehat F_k^{AJ}(t), 
			\label{eq:bias_def} 
		\end{equation} 
		where $1-\widehat S_{\mathrm{KM},k}(t)$ denotes the cause-specific KM 
		risk and 
		$\widehat F_k^{AJ}(t)$ denotes the corresponding AJ 
		estimate of cumulative incidence. 
	\end{definition}
	Thus, $\Biask(t)$ measures the extent to which the KM risk exceeds 
	the corresponding AJ cumulative incidence estimate for event type $k$.
	
	In competing risks settings, cause-specific risks are interdependent 
	because the occurrence of one event type removes the opportunity for the others to occur 
	\cite{allignol2011conditional,latouche2013competing}. 
	Consequently, it is also informative to assess 
	KM overestimation across the competing risks setting as a whole.
	Additionally, in some applications the competing risks system itself is the object of interest.
	For example, in a study about cause of mortality, 
	the primary concern may be the overall impact of competing risks 
	on estimated event probabilities, rather than overestimation for any particular event type.
	This motivates the introduction of a total measure of KM overestimation
	in addition to cause-specific overestimation measures.
	
	\begin{definition}[Empirical Total Overestimation]
	\label{def:total_overestimation}
	We define the {\it empirical total KM overestimation} at time $t$ 
	as the sum of cause-specific overestimations: 
	\begin{equation} 
		\Biast(t) = \sum_{k=1}^{K} \Biask(t). 
		\label{eq:global_bias_def} 
	\end{equation} 
	\end{definition}
	Thus, $\Biast(t)$ measures the total discrepancy between cause-specific KM risks 
	and their corresponding AJ cumulative incidence estimates across all competing event types. 
	Section~\ref{sec:global_diagnostic} 
	derives an exact nonparametric representation of this aggregate quantity 
	using only standard survival analysis estimators, yielding the Total Overestimation Diagnostic.
	
\subsubsection{Decomposition of Empirical Cause-Specific KM Overestimation}
	
	The empirical cause-specific KM overestimation admits a useful decomposition obtained by combining 
	the product limit representation of the KM estimator with the AJ estimator.
	
	\begin{lemma}[Overestimation Decomposition] \label{lem:bias_decomposition} 
		Under independent right censoring, 
		\begin{equation} 
			 \Biask(t) = \sum_{t_i\le t} \left[ \widehat S_{\mathrm{KM},k}(t_{i-1}) - \widehat S_{\mathrm{PL}}(t_{i-1}) \right] \frac{d_k(t_i)}{r(t_i^-)}. \label{eq:bias_decomposition} 
		\end{equation} 
	\end{lemma} 
	\begin{proof}[Derivation] 
	Equation~\eqref{eq:bias_decomposition} follows by combining the discrete-sum representation 
	of the 
	KM complement estimator \cite{pintilie2006competing}, 
	\begin{equation} 
		1-\widehat S_{\mathrm{KM},k}(t) = \sum_{t_i\le t} \widehat S_{\mathrm{KM},k}(t_{i-1}) \frac{d_k(t_i)}{r(t_i^-)}, \label{eq:km_sum_form} 
	\end{equation} 
	with the AJ estimator \eqref{eq:aj_estimator} and substituting both expressions into Equation~\eqref{eq:bias_def}.
	\end{proof}
	
	Equation~\eqref{eq:bias_decomposition} expresses empirical cause-specific KM overestimation as 
	the interaction between a survival function gap generated by competing events and 
	the subsequent occurrence of the event of interest. 
	This representation suggests that the magnitude of KM overestimation may depend 
	not only on competing event frequency but also on the temporal allocation of events. 
	Section~\ref{sec:event_ordering} investigates this possibility through the role of event ordering.
	
\subsubsection{Accumulation of Empirical KM Overestimation} 
	\label{subsec:bias_monotonicity} 

	Equation~\eqref{eq:bias_decomposition} shows that empirical cause-specific KM overestimation accumulates through a sequence of non-negative local contributions.
	The following lemma formalises this accumulation property.
	
	\begin{lemma}[Non-Negativity and Monotonicity of Empirical KM Overestimation] 
		\label{lem:bias_monotonicity} 
		For each cause $k \in \{1,\ldots,K\}$, the empirical cause-specific overestimation function 
		$\Biask(t)$ is a non-negative, monotonically non-decreasing step function of follow-up time. 
		
		Consequently, the empirical total overestimation function 
		$\Biast(t)$ is a non-negative, monotonically non-decreasing step function of follow-up time. 
	\end{lemma} 
	\begin{proof} 
		From Equations~\eqref{eq:pl_estimator} and~\eqref{eq:km_cause_specific}, 
		$d_k(t_i)\le d(t_i)$ at every event time implies that $\widehat S_{\mathrm{KM},k}(t)\ge \widehat S_{\mathrm{PL}}(t)$ for all $t\ge0$.
		Equation~\eqref{eq:bias_decomposition} expresses $\Biask(t)$ as a sum of non-negative terms, 
		since $\widehat S_{\mathrm{KM},k}(t_{i-1})-\widehat S_{\mathrm{PL}}(t_{i-1}) \ge 0$, 
		$d_k(t_i)\ge 0$, and $r(t_i^-)>0$. 
		Therefore, every increment is non-negative and $\widehat B_k(t)$ is a non-negative, monotonically non-decreasing step function of time. 
		Finally, because $\Biast(t) = \sum_{k=1}^K \Biask(t)$ is a finite sum of non-negative, non-decreasing step functions, 
		$\Biast(t)$ inherits these properties.
	\end{proof} 
	
\section{The Role of Event Ordering} 
	\label{sec:event_ordering} 
	
	A common intuition in clinical biostatistics is that KM overestimation is driven primarily by competing event frequency, 
	namely the proportion of individuals experiencing competing events \cite{koller2011competing,austin2017accounting}. 
	However, the decomposition in Equation~\eqref{eq:bias_decomposition} suggests that overestimation 
	may also depend on the temporal allocation of events. 
	This section investigates that possibility and shows that event ordering determines the attainable range of KM overestimation.
		
\subsection{Illustrative Example: Reversing Event Ordering with Fixed Frequencies} 
	\label{subsec:pepe_mori_examples}
	
\begin{figure*}[htb] \centering 
		\begin{minipage}[b]{0.49\textwidth} 
			\includegraphics[width=\textwidth]{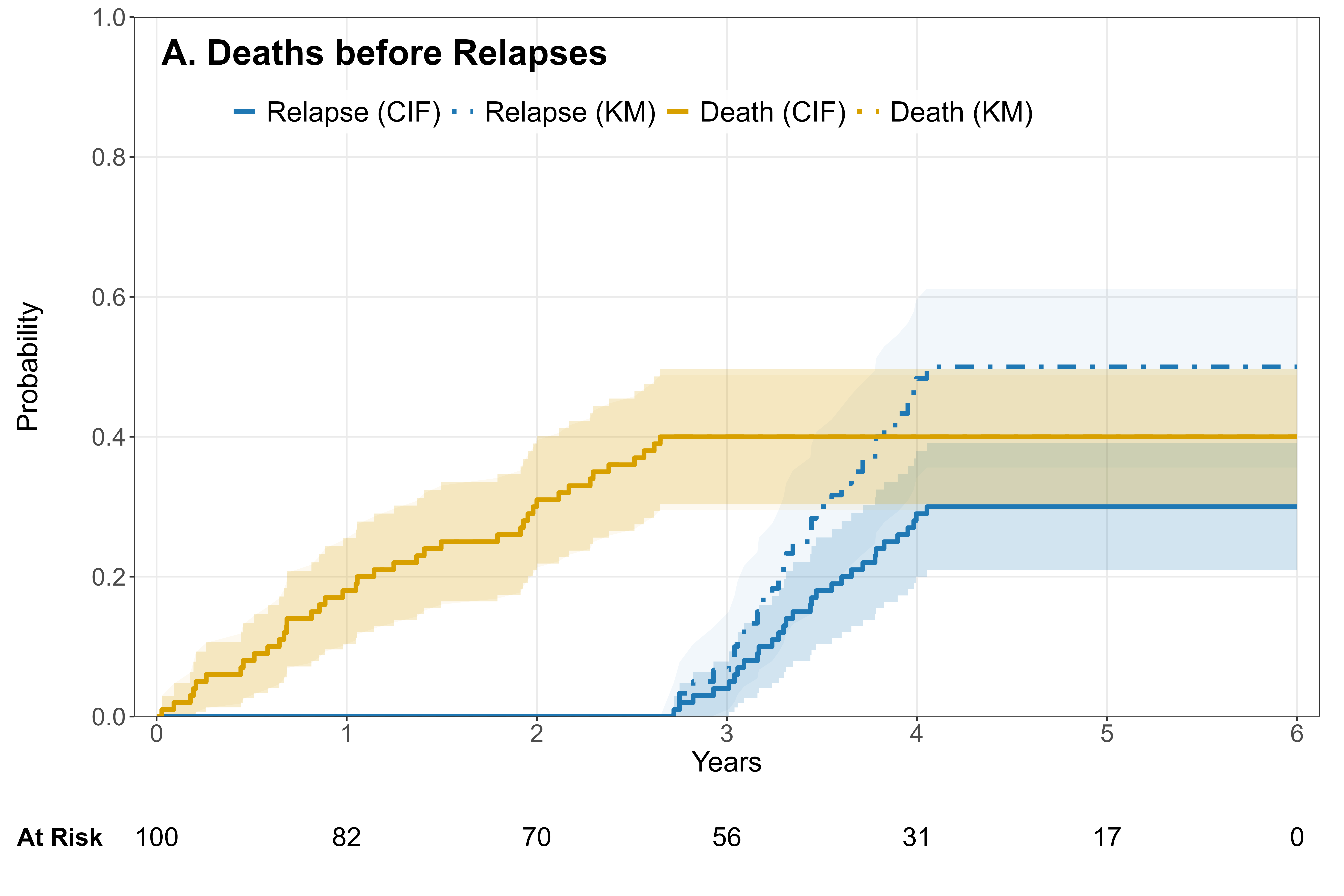} 
		\end{minipage} 
		\hfill 
		\begin{minipage}[b]{0.49\textwidth} 
			\includegraphics[width=\textwidth]{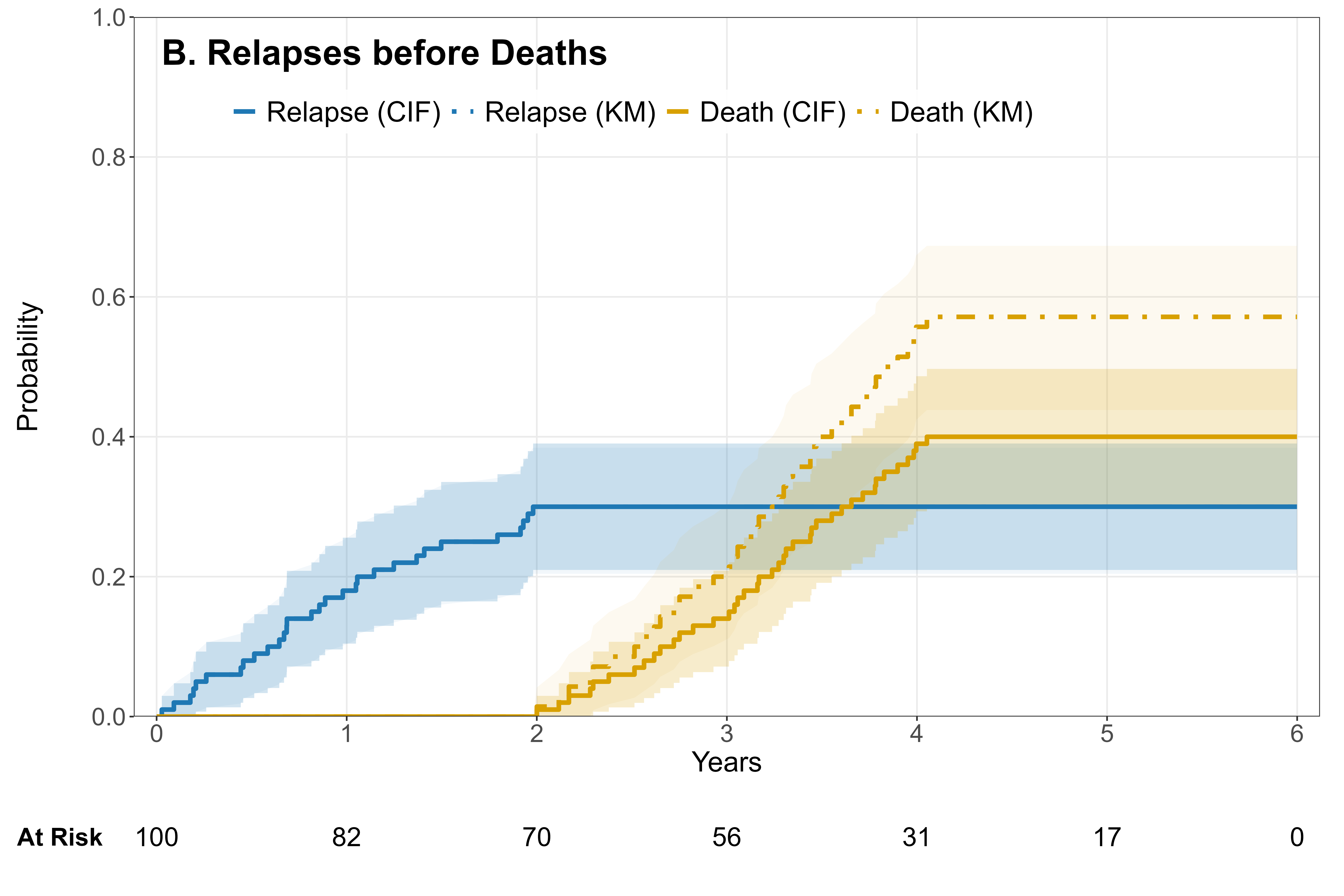} 
		\end{minipage} 
		\caption{ Event ordering as an independent determinant of KM overestimation. 
		Two competing risks settings with identical event frequencies, 
		follow-up duration, and cumulative incidences are shown. 
		Solid curves denote cause-specific cumulative incidence functions (CIFs), 
		dot dashed curves denote KM risks, and shaded bands represent 95\% confidence intervals. 
		Changing only the temporal allocation of events substantially alters KM overestimation, 
		demonstrating that event frequency alone does not determine its magnitude.} 
		\label{fig:pepe_mori} 
	\end{figure*}

	To motivate the formal results that follow, Figure~\ref{fig:pepe_mori} compares 
	two hypothetical cohorts adapted from Pepe and Mori \cite{pepe1993kaplan}. 
	Both cohorts contain $N=100$ individuals, have identical follow-up duration ($t_{\max}=6$ years), 
	identical event frequencies (30 relapses and 40 competing deaths without relapse, 
	followed by 30 observations subject to administrative censoring), 
	and identical cause-specific cumulative incidences at six years: $F_1(6)=0.30$, $F_2(6)=0.40$. 
	The only difference between the two cohorts is the temporal ordering of events. 
	
	\medskip \noindent \textbf{Scenario A (Competing Events First).} 
	All competing deaths occur before any relapse. 
	Consequently, the overall event free survival function $\widehat{S}_{\mathrm{PL}}(t)$ declines before relapses begin, 
	resulting in $1-\widehat{S}_{\mathrm{KM},1}(6)=0.50$, and hence
	 $\widehat{\Bias}_1(6) = \left( 1-\widehat{S}_{\mathrm{KM},1}(6) \right) - F_1(6) = 0.20$.  
	Thus, KM overstates the 6-year relapse probability by 20\%. 
	
	\medskip \noindent \textbf{Scenario B (Relapses First).} 
	The event sequence is reversed while preserving event and censoring frequencies, follow-up duration, and cumulative incidences. 
	All relapses occur before any competing death. 
	Consequently, $1-\widehat{S}_{\mathrm{KM},1}(6)=0.30$, and so 
	$\widehat{\Bias}_1(6)=0$.  
	Reversing the event sequence completely eliminates empirical cause-specific KM overestimation for event type 1
	while preserving event and censoring frequencies, follow-up duration, and cumulative incidences \cite{pepe1993kaplan}. 
	Consequently, the observed difference in overestimation is attributable solely to event ordering. 
	
	This example demonstrates that event frequency alone does not determine KM overestimation and 
	motivates the ordering results developed below.
	
	\begin{remark}[Mechanism of Event Ordering Overestimation] \label{rem:event_ordering_mechanism}
		By Equation~\eqref{eq:bias_decomposition}, empirical cause-specific KM overestimation 
		accumulates through the interaction between a survival function gap 
		generated by competing events and the subsequent occurrence of events of interest. 
		Early competing events enlarge this gap before future events of interest occur, 
		whereas little gap develops when events of interest occur first. 
		Event ordering therefore governs the magnitude of KM overestimation 
		through the alignment between these two processes.
	\end{remark}
	
\subsection{Ordering Bounds  for Empirical Cause-Specific Overestimation} \label{subsec:ordering_bounds}
	
	The preceding example and mechanism  
	suggest that cause-specific KM overestimation may be bounded by extreme ordering configurations. 
	We now characterise these bounds formally. 
	For the ordering results below, event times are assumed to be distinct, $d(t_i)\le1$.
	
	To formalize event ordering, consider a fixed observation skeleton consisting of fixed 
	failure times, fixed censoring times, fixed numbers of events of type $k$, and fixed numbers of competing events.
	 
	Let $\mathcal S_F=(F_1,\ldots,F_m)\in\{E_k,E_c\}^m$ denote the ordered sequence of 
	observed failure labels, where $E_k$ denotes an observed event of type $k$ ($E=k$) and 
	$E_c$ denotes an observed competing event ($E\notin\{0,k\}$).
	\begin{definition}[Failure Inversion Count] 
		The {\it failure inversion count} is defined as 
		\begin{equation} 
			I(\mathcal S_F) = \sum_{1\le i<j\le m} \mathbf 1 \left( F_i=E_k,\, F_j=E_c \right). 
			\label{eq:inversion_count} 
		\end{equation} 
		Thus $I(\mathcal S_F)$ counts the number of instances in which an event of type $k$ occurs before a competing event. 
		The unique zero inversion configuration $\mathcal S_F^\ast = (E_c,\ldots,E_c,E_k,\ldots,E_k)$ places 
		all competing events before all events of type $k$. 
	\end{definition} 
	
	\begin{definition}[Order Index]
		The {\it order index} is defined as
		\begin{equation}
			OI(\mathcal S_F) = 1 - \frac{I(\mathcal S_F)}{n_k n_c}\;,
			\label{eq:OI}
		\end{equation} 
		where $I(\mathcal S_F)$ is the inversion count, $n_k$ is the number of events of interest, 
		and $n_c$ is the number of competing events.
	\end{definition}
	OI values near 0 indicate that events of interest tend to occur before competing events, 
	whereas values near 1 indicate that competing events tend to occur first.
	An OI near 0.50 corresponds to a balanced temporal allocation of event types.
	
\subsubsection{Minimum Cause-Specific Overestimation Order} 
	
	The event ordering that minimises empirical cause-specific KM overestimation corresponds to the following zero overestimation configuration.
	
	\begin{proposition}[Zero Cause-Specific Overestimation Boundary Condition]
		\label{prop:zero_bias_boundary}
		For an event type $k$ with $n_k>0$,
		\[ \Biask(t) = 0, \qquad \forall t \ge 0, \]
		if and only if all competing events occur strictly after the final event of type $k$:
			\[ \mathcal S_F^{\mathbb 0} = (E_k,\ldots,E_k,E_c,\ldots,E_c). \]
		Equivalently, $OI(\mathcal S_F^{\mathbb 0}) = 0$. 
	\end{proposition}
	
	\begin{proof}
	\noindent $(\Leftarrow)$ If all competing events occur strictly after the final event of type $k$, 
		then $\widehat{S}_{\mathrm{KM},k}(t_{i-1}) = \widehat{S}_{\mathrm{PL}}(t_{i-1})$ 
		at every time $t_i$ of cause-specific event of type $k$. 
		Every summand in Equation~\eqref{eq:bias_decomposition} evaluates to zero, 
		implying $\Biask(t) = 0$ for all $t \ge 0$.
		
	\noindent $(\Rightarrow)$ Suppose $\Biask(t) = 0$ for all $t \ge 0$. 
		By Equation~\eqref{eq:bias_decomposition}, $\Biask(t)$ is a sum of non-negative terms:
		\[
			\Biask(t) = \sum_{t_i \le t} \left[ \widehat{S}_{\mathrm{KM},k}(t_{i-1}) - \widehat{S}_{\mathrm{PL}}(t_{i-1}) \right] \frac{d_k(t_i)}{r(t_i^-)}.
		\]
		For this sum to equal zero for all $t \ge 0$, every summand with $d_k(t_i) > 0$ must be zero. Hence
		$\widehat{S}_{\mathrm{KM},k}(t_{i-1}) = \widehat{S}_{\mathrm{PL}}(t_{i-1})$
		at every time $t_i$ of cause-specific event of type $k$.
		Suppose, for contradiction, that at least one competing event precedes an event of type $k$. 
		Let $t_c$ denote the time of the first competing event. 
		At time $t_c$, the competing event reduces overall event free survival to 
		$\widehat{S}_{\mathrm{PL}}(t_c) = \widehat{S}_{\mathrm{PL}}(t_c^-) \left(1 - \frac{1}{r(t_c^-)}\right)$ 
		while leaving KM survival for events of type $k$ unchanged: 
		$\widehat{S}_{\mathrm{KM},k}(t_c) = \widehat{S}_{\mathrm{KM},k}(t_c^-)$. 
		Hence,
		$\widehat{S}_{\mathrm{KM},k}(t_c) > \widehat{S}_{\mathrm{PL}}(t_c)$.
		
		Thereafter, censoring leaves both survival functions unchanged, 
		events of type $k$ multiply both survival functions by the same positive factor, 
		and competing events reduce only $\widehat S_{\mathrm{PL}}$. 
		Consequently, the strict inequality is preserved until the next event of type $k$ occurs.
		
		Let $t_i > t_c$ denote the first subsequent event of type $k$. 
		At time $t_i$, we have $\widehat{S}_{\mathrm{KM},k}(t_{i-1}) > \widehat{S}_{\mathrm{PL}}(t_{i-1})$ and $d_k(t_i) > 0$, 
		yielding a strictly positive summand and implying $\Biask(t) > 0$ for $t \ge t_i$, a contradiction. 
		Therefore, no competing event can precede any event of type $k$. 
		Hence all competing events must occur strictly after the final event of type $k$.
	\end{proof}
	
	\subsubsection{Maximum Cause-Specific Overestimation Order} \label{sec:maxorder}
	
	The zero overestimation boundary in Proposition~\ref{prop:zero_bias_boundary} 
	characterizes one extreme of the event ordering spectrum. 
	We now determine the ordering configuration that maximizes cause-specific KM overestimation. 
		
	The next result characterises the pointwise effect of moving a competing event ahead of an adjacent event of type $k$.
	\begin{lemma}[Transposition of Adjacent Failure Events] 
		\label{lem:km_monotonicity} 
		Let $\mathcal{S}_F$ contain two adjacent failure events $(F_i,F_{i+1})=(E_k,E_c)$ 
		occurring at times $t_a<t_b$, with $c\ge0$ right censored observations between them. 
		Let $\mathcal S'_F$ denote the sequence obtained by interchanging the event labels, 
		so that $(F_i',F_{i+1}') = (E_c,E_k)$. 
		Then the cause-$k$ Kaplan--Meier survival functions satisfy: 
		\begin{enumerate} 
		\item For $t<t_a$, $\widehat S'_{\mathrm{KM},k}(t) =\widehat S_{\mathrm{KM},k}(t)$. 
		\item For $t_a\le t<t_b$, $\widehat S'_{\mathrm{KM},k}(t) >\widehat S_{\mathrm{KM},k}(t)$. 
		\item At $t=t_b$, $\widehat S'_{\mathrm{KM},k}(t_b) <\widehat S_{\mathrm{KM},k}(t_b)$. 
		\item For $t\ge t_b$, $\widehat S'_{\mathrm{KM},k}(t) \le\widehat S_{\mathrm{KM},k}(t)$, 
		with strict inequality unless both KM survival functions have reached zero. 
		\end{enumerate} 
	\end{lemma}
	\begin{proof} 
		Let $r_a=r(t_a^-)$ and $r_b=r(t_b^-)=r_a-1-c$. 
		
		Prior to $t_a$, risk sets and event allocations are identical in both sequences, proving part~1. 
		
		On $[t_a,t_b)$, sequence $\mathcal S_F$ experiences the event $E_k$ at $t_a$, 
		multiplying its cause-$k$ KM survival by $1-1/r_a$. In $\mathcal S'_F$, 
		the event at $t_a$ is competing and therefore does not alter cause-$k$ KM survival. 
		Hence 
		$\widehat S'_{\mathrm{KM},k}(t)$ = $\widehat S_{\mathrm{KM},k}(t_a^-)$ > 
		$\widehat S_{\mathrm{KM},k}(t_a^-) \left(1-\frac1{r_a}\right)$ = 
		$\widehat S_{\mathrm{KM},k}(t)$, $t_a\le t<t_b$, proving part~2. 
		
		At $t_b$, sequence $\mathcal S'_F$ experiences the event $E_k$ against the smaller risk set $r_b<r_a$. 
		Therefore $1-\frac1{r_b}<1-\frac1{r_a}$, and consequently 
		$\widehat S'_{\mathrm{KM},k}(t_b) < \widehat S_{\mathrm{KM},k}(t_b)$, proving part~3. 
		
		After $t_b$, the two sequences have identical remaining event labels, failure times, censoring times, and risk sets. 
		Subsequent cause-$k$ events therefore multiply both KM survival curves by the same non-negative factors, 
		while censoring and competing events leave them unchanged. 
		Hence $\widehat S'_{\mathrm{KM},k}(t) \le \widehat S_{\mathrm{KM},k}(t)$, $t\ge t_b$. 
		The inequality remains strict unless a subsequent common KM factor reduces both curves to zero. 
	\end{proof}
	
	By the preceding lemma, once the transposed pair has occurred, 
	an adjacent swap that removes an inversion increases the corresponding cause-specific KM risk.
	The resulting increase in empirical cause-specific KM overestimation can be quantified explicitly. 
	For a transposition $(E_k,E_c)\rightarrow(E_c,E_k)$ at times $t_a<t_b$, 
	\begin{equation} 
		\Delta \widehat B_k(t_b^+) = 
			\frac{ \widehat B_k(t_a^-)(1+c) + \widehat S_{\mathrm{PL}}(t_a^-) }{ r_a r_b }, 
		\label{eq:local_bias_inflation} 
	\end{equation} 
	where $r_a=r(t_a^-)$ and $r_b=r(t_b^-)=r_a-1-c$. 
	The increment is strictly positive and increases with previously accumulated overestimation, 
	intervening censoring, and remaining event free survival. 
	A derivation is provided in Appendix~\ref{app:local_inflation}.

	The following theorem and corollary characterise the global optimum results arising from Lemma~\ref{lem:km_monotonicity}.
	\begin{theorem}[Global Maximum Ordering of Horizon Cause-Specific KM Risk] 
		\label{thm:km_risk_max} 
		For any fixed observation skeleton, fixed censoring schedule, fixed number of events of type $k$, 
		and fixed number of competing events, 
		the zero inversion block allocation 
			\[ \mathcal S_F^\ast = (E_c,\ldots,E_c,E_k,\ldots,E_k) \] 
		attains the global maximum of the na\"ive cause-specific KM risk at any study horizon 
		$t_{\max}\ge t_m$: 
			$1-\widehat S_{\mathrm{KM},k}(t_{\max})$.  
		If $r(t_i^-)>1$ at every failure time in the fixed observation skeleton, 
		then this maximising allocation is unique. 
	\end{theorem}
	\begin{proof} 
		The result follows by successively removing adjacent inversions. 
		Each transposition cannot decrease cause-specific KM risk at the study horizon, 
		and the process terminates at the zero inversion block allocation $\mathcal S_F^\ast$. 
		
		If $I(\mathcal S_F)>0$, the sequence contains at least one adjacent inversion $(E_k,E_c)$. 
		By Lemma~\ref{lem:km_monotonicity}, swapping this pair cannot decrease the cause-specific KM risk 
		once both failures in the transposed pair have occurred, 
		and therefore cannot decrease the risk at any study horizon $t_{\max}\ge t_m$. 
		The swap also reduces the inversion count by one. 
		Repeated application of adjacent swaps that remove inversions generates a finite sequence 
		\[ \mathcal S_F^{(0)} \longrightarrow \mathcal S_F^{(1)} \longrightarrow \cdots \longrightarrow \mathcal S_F^\ast, \] 
		terminating at the zero inversion block allocation. 
		Because no swap decreases the horizon KM risk, $\mathcal S_F^\ast$ attains the global maximum. 
		
		If $r(t_i^-)>1$ at every failure time, all cause-specific KM factors remain strictly positive. 
		By Lemma~\ref{lem:km_monotonicity}, each adjacent swap that removes an inversion then 
		produces a strict increase in horizon KM risk. 
		Consequently, any sequence with $I(\mathcal S_F)>0$ has strictly smaller horizon KM risk 
		than $\mathcal S_F^\ast$, proving uniqueness. 
	\end{proof}
	\begin{corollary}[Maximum Cause-Specific Overestimation Boundary Condition] 
		\label{cor:theoretical_bias_max} 
		Let $B_k(t_{\max}) = \left[1-\widehat S_{\mathrm{KM},k}(t_{\max})\right] - F_k(t_{\max})$. 
		For fixed $F_k(t_{\max})$, theoretical horizon overestimation is globally maximised 
		by the zero inversion block allocation $\mathcal S_F^\ast$. 
		
		If $r(t_i^-)>1$ at every failure time, 
		this maximum is attained if and only if $\mathcal S_F=\mathcal S_F^\ast$. 
		
		Equivalently, $OI(\mathcal S_F^\ast)=1$.
	\end{corollary}
	\begin{proof} 
		Since $F_k(t_{\max})$ is fixed across all event permutations, 
		maximizing $B_k(t_{\max})$ is equivalent to maximizing $1-\widehat S_{\mathrm{KM},k}(t_{\max})$. 
		Both results follow directly from Theorem~\ref{thm:km_risk_max}. 
	\end{proof}
	
	This section showed that event ordering is not merely associated with cause-specific KM overestimation 
	but determines its attainable range at the study horizon under fixed event frequencies and censoring patterns.
	
	The next section investigates the statistical properties of the empirical cause-specific KM overestimation estimator.
	
\section{Exact Inference for Empirical Cause-Specific Overestimation} \label{sec:statistical_properties}
	
	We now turn our attention to statistical inference.
	A natural question is 
	whether empirical cause-specific KM overestimation requires its own inferential machinery or 
	whether inference can be obtained from existing competing risks methodology.
	
\subsection{AJ Error Equivalence}
	\label{subsec:error_equivalence}

	Recall the empirical cause-specific overestimation estimator defined by Equation~\eqref{eq:bias_def}, 
	    $\Biask(t) = \left( 1 - \widehat S_{\mathrm{KM},k}(t) \right) - \widehat F_k^{\mathrm{AJ}}(t)$,
	and define the corresponding target quantity
	\begin{equation}
	    \Bias_k(t) = \left( 1 - \widehat S_{\mathrm{KM},k}(t) \right) - F_k(t),
	    \label{eq:true_bias_def}
	\end{equation}
	where $F_k(t)$ denotes the true cause-specific cumulative incidence function. 
	The quantity $\Bias_k(t)$ 
	represents the true cause-specific KM overestimation at time $t$.
	
	\begin{theorem}[AJ Error Equivalence]
		\label{thm:error_equivalence}
		For any event type $k \in \{1, \dots, K\}$ and any follow-up time $t \ge 0$,
		\begin{equation}
		    \Biask(t) - \Bias_k(t) = F_k(t) - \widehat F_k^{\mathrm{AJ}}(t).
		    \label{eq:error_equivalence}
		\end{equation}
	\end{theorem}
	
	\begin{proof}
	Subtracting Equation~\eqref{eq:true_bias_def} from Equation~\eqref{eq:bias_def} gives
	\begin{align*}
	    \Biask(t) - \Bias_k(t) &= 
		    \Big[ \left(1 - \widehat S_{\mathrm{KM},k}(t)\right) - \widehat F_k^{\mathrm{AJ}}(t) \Big] - 
		    \Big[ \left(1 - \widehat S_{\mathrm{KM},k}(t)\right) - F_k(t) \Big] \\
	    &= F_k(t) - \widehat F_k^{\mathrm{AJ}}(t),
	\end{align*}
	which proves Equation~\eqref{eq:error_equivalence}.
	\end{proof}
	
\subsection{Inferential Consequences for Cause-Specific Overestimation} \label{subsec:conseqs_AJ_thm}

	The AJ Error Equivalence theorem immediately yields the following consistency and exact variance inheritance results.
	
	\begin{corollary}[Cause-Specific Overestimation Consistency Inheritance] 
		\label{cor:consistency}
		Under independent right censoring and a non-vanishing risk set,	
		the empirical cause-specific overestimation estimator $\Biask$
		is uniformly consistent for $\Bias_k$:
		\begin{equation}
		    \| \Biask - \Bias_k \| _\infty \xrightarrow{\Prob} 0.
		    \label{eq:consistency}
		\end{equation}
	\end{corollary}
	
	\begin{proof}
		Under independent right censoring
		and a non-vanishing risk set, the AJ estimator is uniformly consistent, 
		$\| F_k-\widehat F_k^{AJ} \|_\infty \xrightarrow{\Prob}0$ 
		\cite{aalen1978empirical,kalbfleisch2002statistical,geskus2016data}. 
		By Theorem~\ref{thm:error_equivalence},
		$\| \Biask - \Bias_k \| _\infty = \| F_k -\widehat F_k^{AJ} \|_\infty \xrightarrow{\Prob}0$, which gives the result.
	\end{proof}
	
	\begin{corollary}[Cause-Specific Overestimation Variance Inheritance] 
		\label{cor:variance}
		For any event type $k \in \{1, \dots, K\}$ and follow-up time $t \ge 0$,
		\begin{equation}
		    \Var\!\left( \Biask(t) - \Bias_k(t) \right) = \Var\!\left( \widehat F_k^{\mathrm{AJ}}(t) \right).
		    \label{eq:variance_identity}
		\end{equation}
	\end{corollary}
	
	\begin{proof}
		Taking variances on both sides of Equation~\eqref{eq:error_equivalence} gives
		$\Var\!\left( \Biask(t) - \Bias_k(t) \right) = \Var\!\left( F_k(t) - \widehat F_k^{\mathrm{AJ}}(t) \right)$.
		
		Since $F_k(t)$ is non-random,
		$\Var\!\left( F_k(t) - \widehat F_k^{\mathrm{AJ}}(t) \right) = \Var\!\left( \widehat F_k^{\mathrm{AJ}}(t) \right)$,
		which proves Equation~\eqref{eq:variance_identity}.
	\end{proof}

	Theorem~\ref{thm:error_equivalence} and Corollaries~\ref{cor:consistency}--\ref{cor:variance} 
	show that empirical cause-specific KM overestimation
	inherits the inferential framework of the AJ estimator. 
	As a result, standard AJ confidence intervals, variance estimators, 
	and large sample results apply directly to cause-specific KM overestimation 
	without modification or additional asymptotic theory.
	Inference reduces directly to inference for the corresponding AJ estimator.
	
\section{Exact Diagnostic for Empirical Total KM Overestimation} 
	\label{sec:global_diagnostic} 
	
	We now extend the previous cause-specific framework to the competing risks setting as a whole.
	This section addresses three related questions. 
	
	How can total KM overestimation be quantified? \\
	How can uncertainty in total KM overestimation be quantified?\\
	Is total KM overestimation an unavoidable feature of competing risks settings?
	
\begin{figure*}[htbp] \centering 
	 	\begin{minipage}[b]{0.49\textwidth} 
			\includegraphics[width=\textwidth]{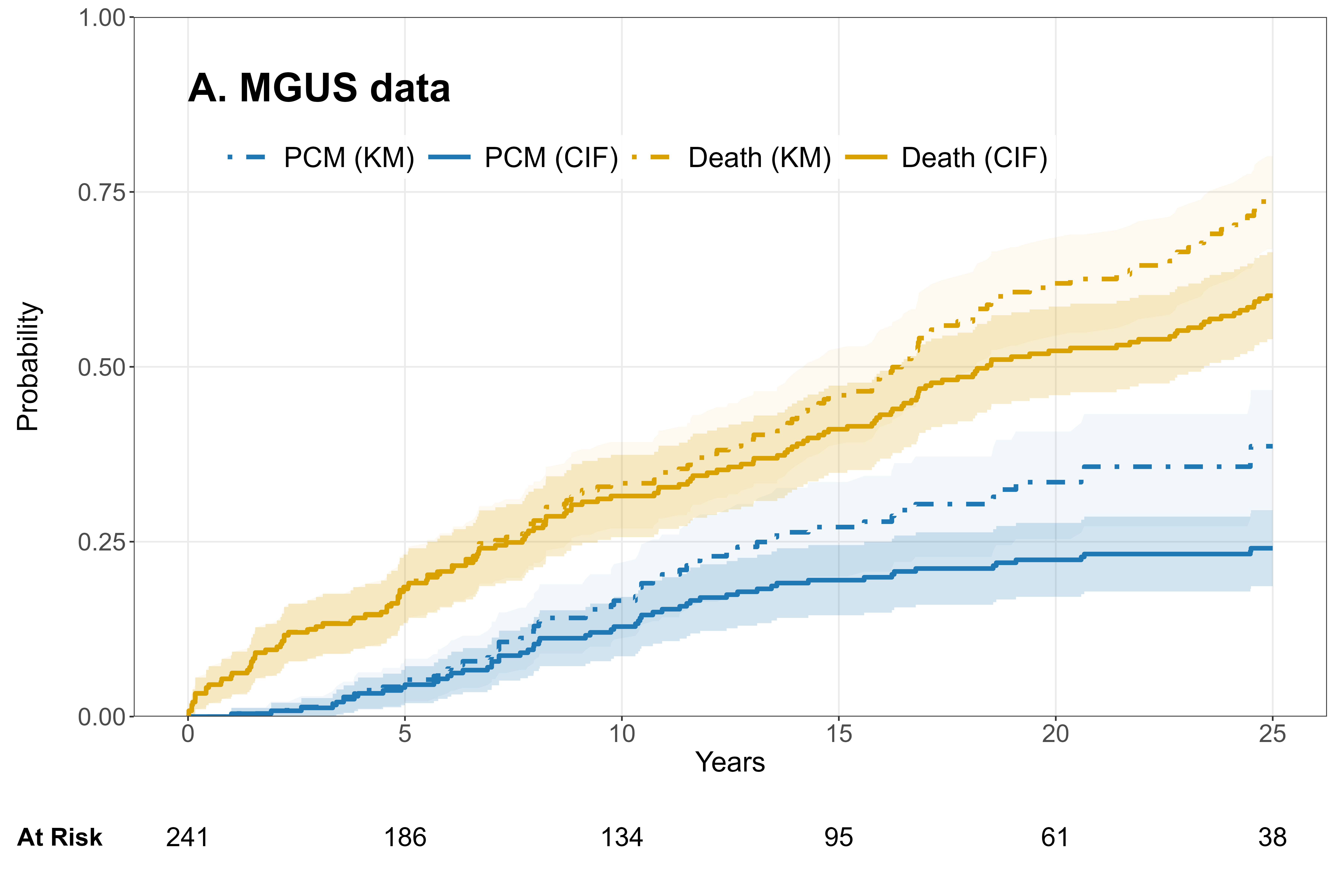} 
		\end{minipage} 
		\hfill 
		\begin{minipage}[b]{0.49\textwidth} 
			\includegraphics[width=\textwidth]{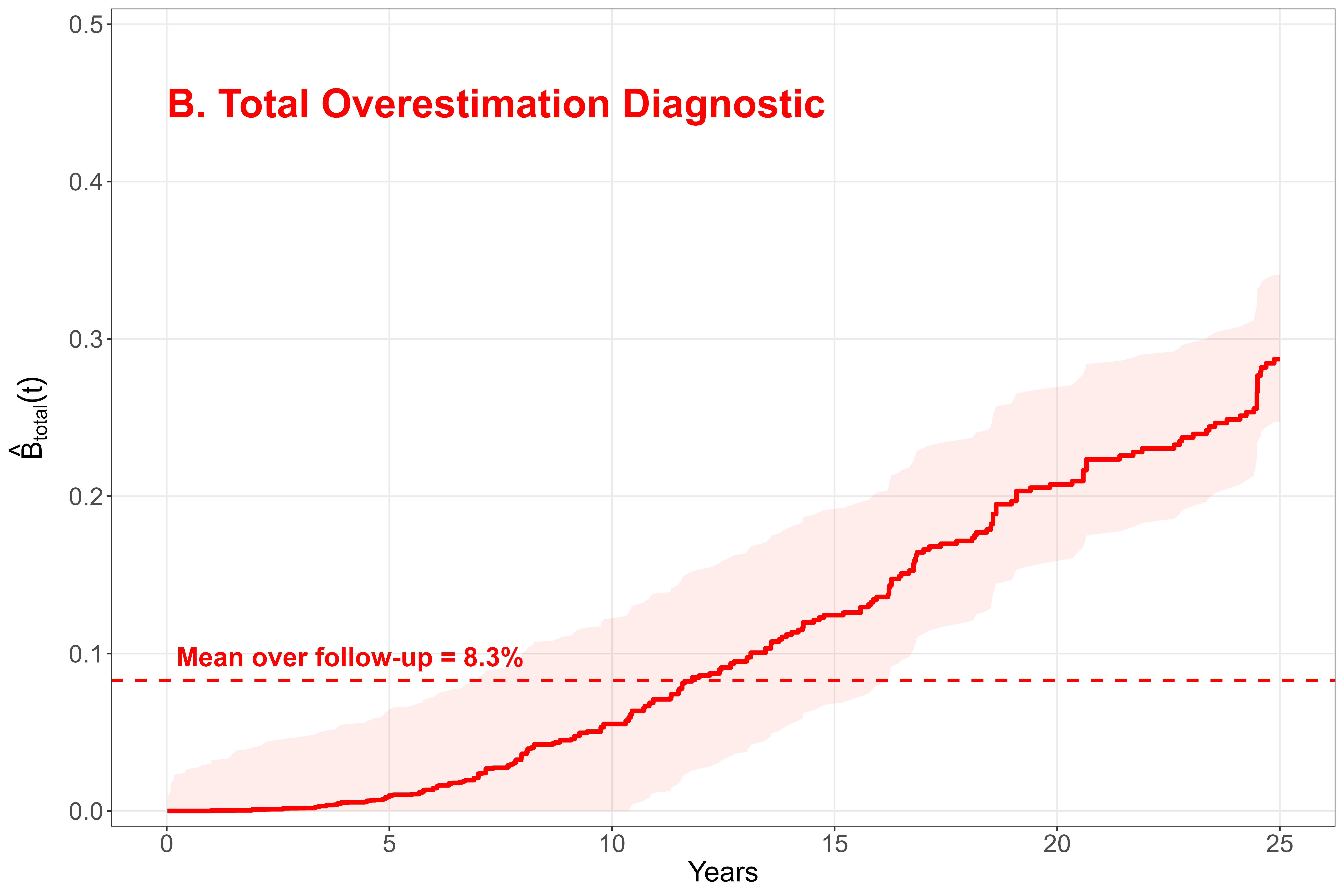} 
		\end{minipage} 
		  \caption{MGUS competing risks overestimation audit. 
		  Panel A compares cause-specific KM risks with the corresponding Aalen--Johansen cumulative incidence functions. 
		  Panel B displays the Total Overestimation Diagnostic (TOD) together with 95\% confidence intervals. 
		  The figure illustrates that total overestimation and its uncertainty 
		  can be quantified directly using routine survival analysis output.} 
		  \label{fig:mgus} 
	  \end{figure*}

	The resulting {\it Total Overestimation Diagnostic (TOD)} 
	quantifies empirical total KM overestimation 
	across all competing event types using only standard nonparametric survival estimators.
	Figure~\ref{fig:mgus}  illustrates TOD in the MGUS cohort. 
	Full details on this empirical evaluation are provided in Section~\ref{subsec:mgus}.
	
\subsection{Exact Quantification of Total Overestimation} 
	\label{subsec:global_diagnostic_identity}
	
	The first question is straightforward: How large is total KM overestimation?
		
	The following theorem provides a closed-form quantification 
	of empirical total KM overestimation through the Total Overestimation Diagnostic (TOD).
	
	\begin{theorem}[Closed-Form Total Overestimation Diagnostic] 
		\label{thm:global_diagnostic}
	
		Under independent right censoring and a non-vanishing risk set, 
		the empirical total KM overestimation $\Biast(t)$ can be calculated as
		\begin{equation}
		    \boxed{
		    		TOD(t) = \widehat S_{\mathrm{PL}}(t)  - \sum_{k=1}^{K} \widehat S_{\mathrm{KM},k}(t) + (K - 1).
		    	}
		    \label{eq:global_diagnostic_identity}
		\end{equation}
	 \end{theorem}
	
	Theorem~\ref{thm:global_diagnostic} provides 
	an exact measure of total KM overestimation using only 
	standard overall product limit and cause-specific KM estimators. 
	In particular, total KM overestimation can be quantified directly from routine survival analysis output 
	without requiring cumulative incidence calculations.
	
	\begin{proof}
		Substituting Equation~\eqref{eq:bias_def} into $\Biast(t) = \sum_{k=1}^{K} \Biask(t)$ 
		and using the risk decomposition identity for the AJ estimator \cite{kalbfleisch2002statistical,geskus2016data},
		$\sum_{k=1}^{K} \widehat F_k^{\mathrm{AJ}}(t) = 1 - \widehat S_{\mathrm{PL}}(t)$, yields
		Equation~\eqref{eq:global_diagnostic_identity}.
	\end{proof}	

	\begin{remark}[TOD for Two Competing Risks]	
		For the common case of two competing event types ($K=2$), 
		Equation~\eqref{eq:global_diagnostic_identity} for quantifying $\Biast(t)$ simplifies to 
			\[ TOD(t) = \widehat S_{\mathrm{PL}}(t) - \widehat S_{\mathrm{KM},1}(t) - \widehat S_{\mathrm{KM},2}(t) +1. \] 
		This representation allows total KM overestimation to be calculated directly from three standard survival curves.
	\end{remark}	

\subsection{Exact Inference for Total Overestimation}
	\label{subsec:global_error_equivalence}

	The second question concerns statistical inference: 
	How can uncertainty in total overestimation be quantified?

	The following theorem is the key inferential result of this paper. 
	It determines an exact, finite sample error equivalence identity 
	between the total overestimation diagnostic and the overall product limit estimator.

\subsubsection{Product Limit Error Equivalence}

	The Total Overestimation Diagnostic (TOD) introduced above possesses a remarkable inferential property. 
	Define the target total overestimation quantity as
	\begin{equation}
	    \Bias_{\mathrm{total}}(t) = \sum_{k=1}^{K} \Bias_k(t) 
	    \overset{(\ref{eq:true_bias_def},\ref{eq:risk_decomposition})}{=} 
	    \left( 1 - S(t) \right) - \sum_{k=1}^{K} \widehat S_{\mathrm{KM},k}(t) + (K - 1),
	    \label{eq:true_global_bias}
	\end{equation}
	where $S(t) = \Prob(T > t)$ denotes the true overall event free survival function. 
	The quantity $\Bias_{\mathrm{total}}(t)$ represents the total KM overestimation 
	that would be observed if the true overall survival function $S(t)$ were known.
	
	\begin{theorem}[Product Limit Error Equivalence]
		\label{thm:global_error_equivalence}
		For any follow-up time $t \ge 0$,
		\begin{equation}
			\boxed{
			    \Biast(t) - \Bias_{\mathrm{total}}(t) = \widehat S_{\mathrm{PL}}(t) - S(t).
			}
			\label{eq:global_error_equivalence}
		\end{equation}
	\end{theorem}
	
	\begin{proof}
		Aggregating the cause-specific error-equivalence identity of Theorem~\ref{thm:error_equivalence} 
		over all event types $k = 1, \dots, K$ gives
		\[  \Biast(t) - \Bias_{\mathrm{total}}(t) = \sum_{k=1}^{K} F_k(t) - \sum_{k=1}^{K} \widehat F_k^{\mathrm{AJ}}(t). \]
		Using the theoretical and empirical risk decomposition identities \cite{kalbfleisch2002statistical,geskus2016data},
		$\sum_{k=1}^{K} F_k(t) \overset{\eqref{eq:risk_decomposition}}{=} 1 - S(t)$ 
		 and
		 $\sum_{k=1}^{K} \widehat F_k^{\mathrm{AJ}}(t) = 1 - \widehat S_{\mathrm{PL}}(t)$,
		gives Equation~\eqref{eq:global_error_equivalence}.
	\end{proof}
	
	Theorem~\ref{thm:global_error_equivalence} reveals a complete simplification of the total overestimation inferential problem. 
	Inference for total KM overestimation is not merely 
	asymptotically equivalent to inference for overall survival; 
	the two error processes are algebraically identical for every sample and every follow-up time.
	
\subsubsection{Inferential Consequences for the Total Overestimation Diagnostic}
	\label{subsec:global_inference}
	
	The Product Limit Error Equivalence theorem immediately yields 
	consistency and exact variance inheritance for the total overestimation diagnostic.
	
	\begin{corollary}[Total Overestimation Consistency Inheritance]
		\label{cor:global_consistency}
		Under independent right censoring and a non-vanishing risk set,
		the empirical total overestimation estimator $\Biast$
		is uniformly consistent for $\Bias_{\mathrm{total}}$:
		\begin{equation}
		    \| \Biast - \Bias_{\mathrm{total}} \|_\infty \xrightarrow{\Prob} 0.
		    \label{eq:global_consistency}
		\end{equation}
	\end{corollary}
	
	\begin{proof}
		Under independent right censoring (assumption of Section~\ref{subsec:background}),
		the product limit estimator for overall event free survival is uniformly consistent \cite{aalen1978empirical,geskus2016data},
		and so $\| \widehat S_{\mathrm{PL}} - S \|_\infty \xrightarrow{\Prob} 0$.
	
		Applying Theorem~\ref{thm:global_error_equivalence}, we get:
		$\| \Biast - \Bias_{\mathrm{total}} \|_\infty = \| \widehat S_{\mathrm{PL}} - S \|_\infty \xrightarrow{\Prob} 0$.
	\end{proof}
	
	\begin{corollary}[Total Overestimation Variance Inheritance]
		\label{cor:global_variance}
		For any follow-up time $t \ge 0$,
		\begin{equation}
		    \Var\!\left( \Biast(t) - \Bias_{\mathrm{total}}(t) \right) = \Var\!\left( \widehat S_{\mathrm{PL}}(t) \right).
		    \label{eq:global_variance}
		\end{equation}
	\end{corollary}
	
	\begin{proof}
		Taking variances on both sides of Equation~\eqref{eq:global_error_equivalence} gives
		$\Var\!\left( \Biast(t) - \Bias_{\mathrm{total}}(t) \right) = \Var\!\left( \widehat S_{\mathrm{PL}}(t) - S(t) \right)$.
		
		Since $S(t)$ is non-random,
		the previous equality can be simplified as Equation~\eqref{eq:global_variance}.
	\end{proof}
	
	Consequently, all standard product limit inferential procedures \cite{kalbfleisch2002statistical,geskus2016data} apply directly to TOD.
	
\subsection{Structural Properties of Total Overestimation} 
	\label{subsec:global_structure} 
	
	A fundamental question is whether total overestimation can ever be absent.
	If total overestimation is merely a possible consequence of competing risks, 
	its practical importance may depend on the specific dataset under study. 
	The following result establishes a stronger conclusion.	
	
	\begin{theorem}[Structural Positivity of Total Overestimation at Horizon] 
		\label{thm:tod_strict_positivity} 
		Consider a competing risks setting with $K \ge 2$ event types.
		Let $t_{\max}$ denote the study horizon, 
		with $t_{\max} \ge t_m$, where $t_m$ is the final observed failure time.
		If at least two distinct event types occur during follow-up, 
		then the Total Overestimation Diagnostic is strictly positive at end of follow-up: 
		\begin{equation} 
			\Biast(t_{\max})  
				> 0. 
			\label{eq:tod_strict_positivity} 
		\end{equation} 
	\end{theorem} 
	\begin{proof} 
		Without loss of generality, assume that the first occurring failure event 
		belongs to event type~1 and occurs at time $t_1$, 
		and let $t_c > t_1$ denote the first occurrence of a different realised event type, labelled event type~2.  
		
		At time $t_1$, the overall product limit survival estimator decreases, 
		$\widehat S_{\mathrm{PL}}(t_1) = 1-\frac{1}{r(t_1^-)} < 1$, 
		whereas the event type~2 Kaplan--Meier survival estimator remains unchanged, 
		$\widehat S_{\mathrm{KM},2}(t_1) = 1$.
		Consequently, 
		$\widehat S_{\mathrm{KM},2}(t_1) - \widehat S_{\mathrm{PL}}(t_1) = \frac{1}{r(t_1^-)} > 0$. 
		
		At time $t_c^{-}$, this survival gap remains strictly positive. 
		Since the event at time $t_c$ is of type~2,
		Equation~\eqref{eq:bias_decomposition} implies that 
		the corresponding increment in $\widehat B_2(t_c)$ is strictly positive: $\widehat B_2(t_c)>0$.  
		
		By Lemma~\ref{lem:bias_monotonicity}, 
		$\widehat B_2(t_{\max}) \ge \widehat B_2(t_c) > 0$.  
		
		Moreover, empirical overestimation is non-negative for every cause, 
		$\widehat B_k(t_{\max})\ge0$, $k = 1, \dots, K$. 
	
		Finally, $\Biast(t_{\max}) = \sum_{k=1}^K \widehat B_k(t_{\max}) \ge \widehat B_2(t_{\max}) \geq \widehat B_2(t_c) > 0.$  
	\end{proof} 

	\begin{remark}[Total Overestimation is Structurally Unavoidable] 
		\label{cor:tod_nonvanishing} 
		In any competing risks dataset containing at least two distinct realised event types, 
		the Total Overestimation Diagnostic cannot vanish at study horizon.
		Consequently, total Kaplan--Meier overestimation is an 
		unavoidable structural feature of the observed competing risks setting whenever at least two event types occur. 
	\end{remark} 
	
	The practical significance of Section~\ref{sec:global_diagnostic} is threefold.
	TOD provides an exact diagnostic for total KM overestimation using only standard survival analysis estimators. 
	Product Limit Error Equivalence reduces uncertainty quantification for TOD to ordinary overall survival estimation. 
	Structural positivity establishes that total KM overestimation
	is unavoidable whenever multiple event types are realised. 
	Consequently, total overestimation is not a phenomenon that must first be detected; 
	it is always present and it must be quantified and reported.
	
 	Together, event ordering theory, TOD, Product Limit Error Equivalence, and Structural Positivity 
 	imply that KM overestimation is an unavoidable,
 	measurable, inferable, and reportable feature of competing risks settings. 
 	
 	The next section evaluates these theoretical results using benchmark datasets and simulation studies 
 	before turning to their implications for assessing competing risks relevance.

\section{Empirical Evaluation} \label{sec:empirical_evaluation} 

\subsection{Benchmark Competing Risk Datasets \\ 
\textit{(Frequency alone is insufficient)}} 
	\label{subsec:benchmark} 
	
	To evaluate the empirical behaviour of the proposed framework, 
	we analysed six benchmark datasets widely used in survival and competing risks research. 
	The datasets span plasma cell disorders, hepatology, oncology, transplantation, and acute leukaemia.
	
	The benchmark panel includes 
	the MGUS cohort, where progression to plasma cell malignancy 
	competes with death without prior progression \cite{kyle1993benign,kyle2002review}; 
	the PBC and UDCA cohorts, where death 
	competes with liver transplantation \cite{therneau2000modeling,lindor1994ursodeoxycholic}; 
	the MELANOMA cohort, where melanoma-specific mortality 
	competes with death from other causes \cite{andersen1993counting}; 
	the TRANSPLANT cohort, where liver transplantation 
	competes with waiting list mortality \cite{kim2006deaths}; 
	and the MYELOID cohort, where relapse 
	competes with death before relapse \cite{lerademacher2018multistate}.
	
	These datasets were chosen because they are widely used benchmark examples 
	in the survival, multi-state and competing risks literature. 
	For comparability, each dataset was administratively censored at a clinically meaningful fixed horizon, 
	and event definitions were harmonised to define a single event of interest and a single competing event.

	Table~\ref{tab:benchmark_datasets} demonstrates considerable 
	heterogeneity in KM overestimation across these widely used benchmark datasets. 
	Cause-specific KM overestimation ranged from 0.4\% in the UDCA cohort to 14.6\% in MGUS, 
	while TOD ranged from 0.6\% to 28.7\%. 
	
	\begin{table*}[htbp] \centering 
		\caption{ Benchmark competing risks datasets illustrating the practical heterogeneity of Kaplan--Meier (KM) overestimation across clinical domains. 
		Despite similar competing event frequencies in some datasets, 
		observed KM overestimation varies substantially. 
		Competing Events denotes the proportion of competing events among all observed events;
		Order Index (OI) is defined by Equation~\eqref{eq:OI}; 
		KM--CIF denotes empirical cause-specific KM overestimation; 
		and TOD denotes the Total Overestimation Diagnostic introduced in Equation~\eqref{eq:global_diagnostic_identity}. }
		\label{tab:benchmark_datasets} 
		\footnotesize \begin{tabular*}{\textwidth}{@{\extracolsep{\fill}}l c c c c cc c c@{}} 
		\toprule 
		Dataset & N & Censoring & Competing Events & Order Index & KM Risk & CIF & KM--CIF & TOD \\ 
		\midrule 
		PBC               & 418 & 0.567 & 0.138 & 0.392 & 0.558 & 0.527 & 0.031 & 0.067 \\ 
		MELANOMA & 205 & 0.654 & 0.197 & 0.533 & 0.355 & 0.339 & 0.016 & 0.049 \\ 
		UDCA           & 170 & 0.894 & 0.389 & 0.597 & 0.115 & 0.111 & 0.004 & 0.006 \\ 
		MYELOID      & 646 & 0.418 & 0.402 & 0.604 & 0.447 & 0.369 & 0.077 & 0.137 \\ 
		MGUS           & 241 & 0.158 & 0.714 & 0.502 & 0.386 & 0.241 & 0.146 & 0.287 \\ 
		TRANSPLANT & 815 & 0.150 & 0.909 & 0.412 & 0.161 & 0.081 & 0.081 & 0.142 \\ 
		\bottomrule 
		\end{tabular*} 
	\end{table*}	
	
	These results indicate that competing event frequency alone is insufficient to characterise the magnitude of overestimation.
	For example, the MYELOID and UDCA cohorts exhibited similar 
	competing event frequencies (40.2\% and 38.9\%, respectively), 
	yet differed almost twenty fold in observed KM overestimation (7.7\% versus 0.4\%).

\subsection{Empirical Application: MGUS Cohort \\ 
\textit{(TOD is measurable and reportable)}} 
	\label{subsec:mgus} 
	
	To evaluate the empirical behaviour of the proposed framework, 
	we analysed the classical Monoclonal Gammopathy of Undetermined Significance (MGUS) cohort, 
	a widely used competing risks benchmark dataset describing 
	progression to plasma cell malignancy or death of 241 patients 
	\cite{kyle1993benign,kyle2002review,geskus2016data}. 
	For comparability, follow-up was administratively censored at 25 years. 
	At this horizon, 58 patients (24.1\%) had progressed to plasma cell malignancy, 
	145 (60.2\%) had died before progression, and 38 (15.8\%) remained event free.
	
	Figure~\ref{fig:mgus} presents both the conventional KM versus AJ comparison 
	and the corresponding Total Overestimation Diagnostic. 
	Across the observed 25 year follow-up period, 
	mean progression KM overestimation was approximately 4.8 percentage points,  
	mean competing mortality KM overestimation was approximately 3.5 percentage points, and  
	mean total KM overestimation was approximately 8.3 percentage points. 
	These means provide descriptive summaries across follow-up. 
	At the prespecified 25-year horizon, TOD was 0.287, 
	indicating that the sum of the cause-specific KM risks exceeded the 
	corresponding cumulative incidences by 28.7 percentage points (95\% CI, 0.247--0.341).
	
	To verify the closed-form identity numerically, we defined the diagnostic residual
	$R(t)=\Biast(t)-\sum_{k=1}^{K}\Biask(t)$. 
	As predicted by Theorem~\ref{thm:global_diagnostic},
	TOD recovered 
	the observed total overestimation across the entire follow-up period, 
	to numerical precision. 

	Confidence intervals for TOD were obtained directly from standard 
	product limit confidence limits through Product Limit Error Equivalence. 
	The resulting confidence bands behaved as predicted, 
	confirming that uncertainty quantification for total overestimation 
	requires no competing risks specific inferential machinery.	

	The MGUS cohort illustrates how TOD converts that overestimation 
	into a reportable quantity with formal statistical uncertainty. 
	Together with Structural Positivity (Theorem~\ref{thm:tod_strict_positivity}), 
	this shifts the practical question from whether total overestimation exists 
	to how large it is.
	TOD thus provides a practical audit metric for KM overestimation 
	in routine competing risks analyses.
	
\subsection{Frequency, Ordering, and the Magnitude of KM Overestimation\\
 \textit{(Ordering matters)}} 
 	\label{subsec:phase4_simulation} 
	
	The event ordering theory developed in Section~\ref{sec:event_ordering} predicts that 
	the magnitude of KM overestimation depends not only on competing event frequency 
	but also on the temporal ordering of competing events and events of interest. 
	To investigate the practical implications of this result, 
	we conducted a large scale simulation study spanning
	a broad range of competing risk settings. 
	Simulated datasets were generated  
	under varying competing event frequencies and ordering structures, 
	using a fixed sample size of N=500 and approximately 20\% independent right censoring. 
	Complete simulation details are provided in Appendix~\ref{app:phase4_design},
	while calibration and quality control diagnostics are reported in Appendix~\ref{app:phase4_calibration}.	
	
	\begin{figure*}[htbp] \centering 
		\includegraphics[width=0.49\textwidth]{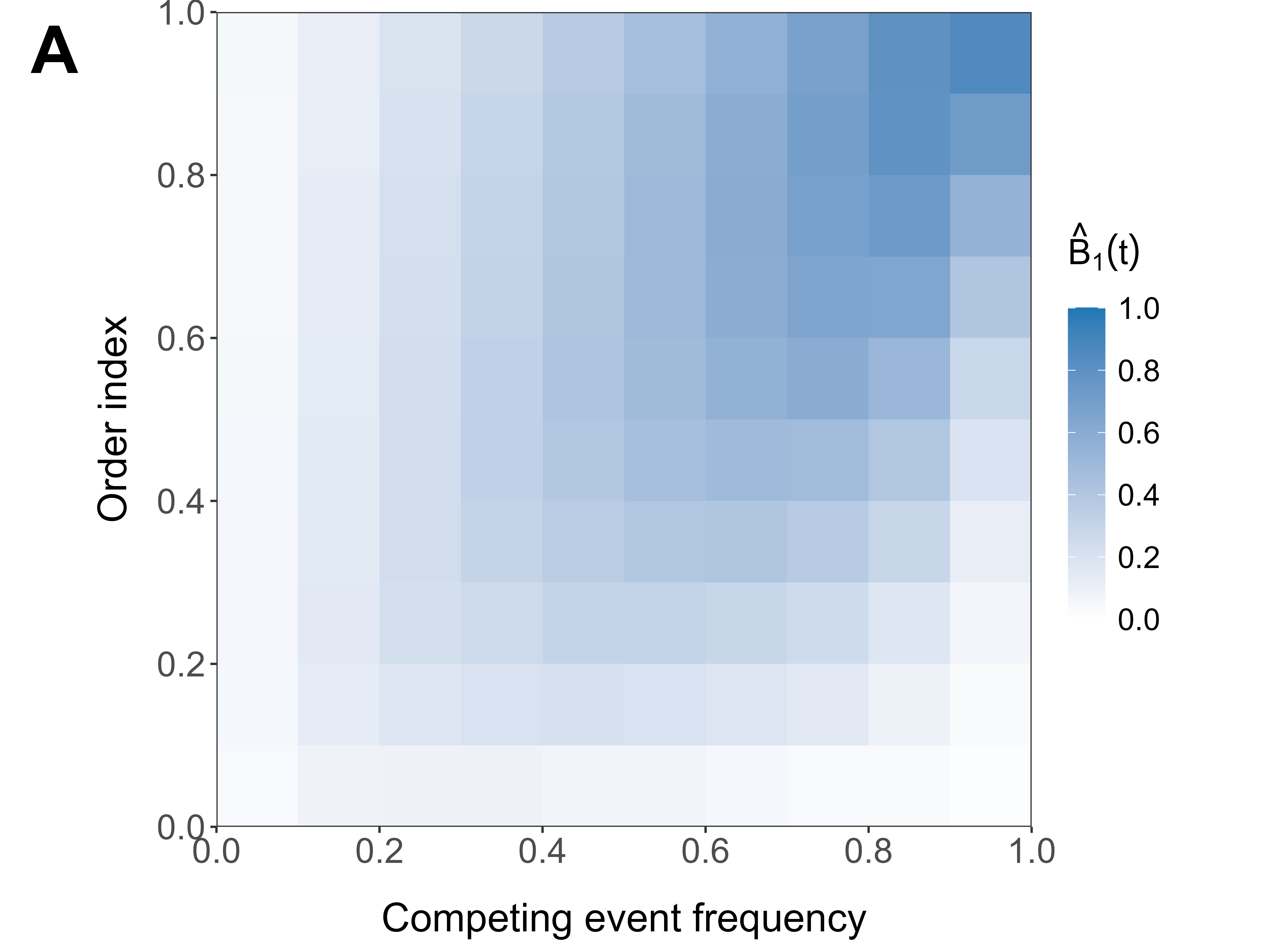}		
		\includegraphics[width=0.49\textwidth]{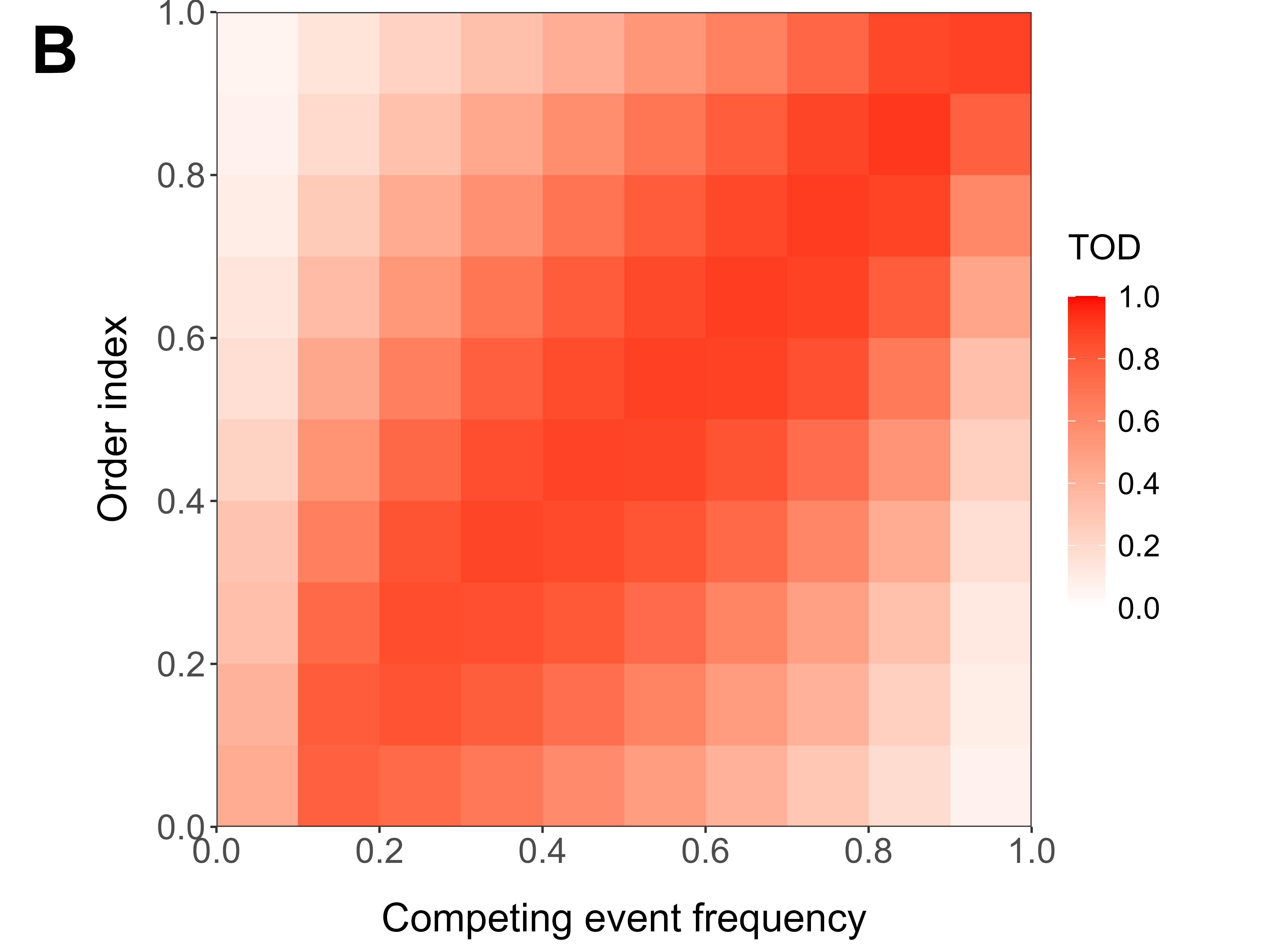}
		\caption{ Impact of competing event frequency and event ordering on KM overestimation. 
		Panel A displays mean cause-specific KM overestimation and
		Panel B the mean Total Overestimation Diagnostic (TOD). 
		The vertical axis corresponds to the Order Index (OI); 
		higher OI values indicate earlier competing events. 
		Under identical competing event frequencies, 
		different OI values produce markedly different levels of overestimation.}
		 \label{fig:tod_heatmaps} 		
	 \end{figure*}

	Figure~\ref{fig:tod_heatmaps} summarises 
	mean cause-specific KM overestimation for the event of interest (Panel~A) and 
	mean Total Overestimation Diagnostic (TOD; Panel~B) under representative simulation settings. 
	The vertical axis corresponds to the Order Index (OI) defined by Equation~\eqref{eq:OI}, 
	with lower values indicating that events of interest tend to occur earlier than competing events 
	and higher values indicating the opposite.
	
	Several findings emerge. 
	Most importantly, 
	event ordering materially alters both cause-specific and total overestimation even when competing event frequency is held fixed. 	
	The influence of ordering becomes increasingly pronounced as competing event frequency increases. 
	At low competing event frequencies, differences between ordering configurations are comparatively modest. 
	At moderate and high competing event frequencies, however, markedly different levels of overestimation arise 
	under identical competing event frequencies but different Order Index values.
	
	Cause-specific and total overestimation also exhibit distinct behaviour. 
	Configurations producing the largest overestimation for the event of interest 
	do not necessarily produce the largest TOD values. 
	This observation suggests that mechanisms governing 
	extremal cause-specific and total overestimation may differ, 
	highlighting the distinction between cause-specific and total measures of overestimation.
	
	The simulations provide empirical support 
	for the event ordering theory developed in Section~\ref{sec:event_ordering}. 
	They further 
	show that the practical importance of event ordering increases as competing event frequency increases,
	with markedly different levels of KM overestimation arising 
	under identical competing event frequencies but different temporal event allocations.
	These results confirm that temporal event allocation is an independent determinant of both cause-specific and total KM overestimation.	
	
\subsection{Censoring Stress Test \\
\textit{(Product Limit Error Equivalence works)}} 
	\label{subsec:censoring_stress} 
	
\begin{figure*}[htbp] \centering \includegraphics[width=\textwidth]{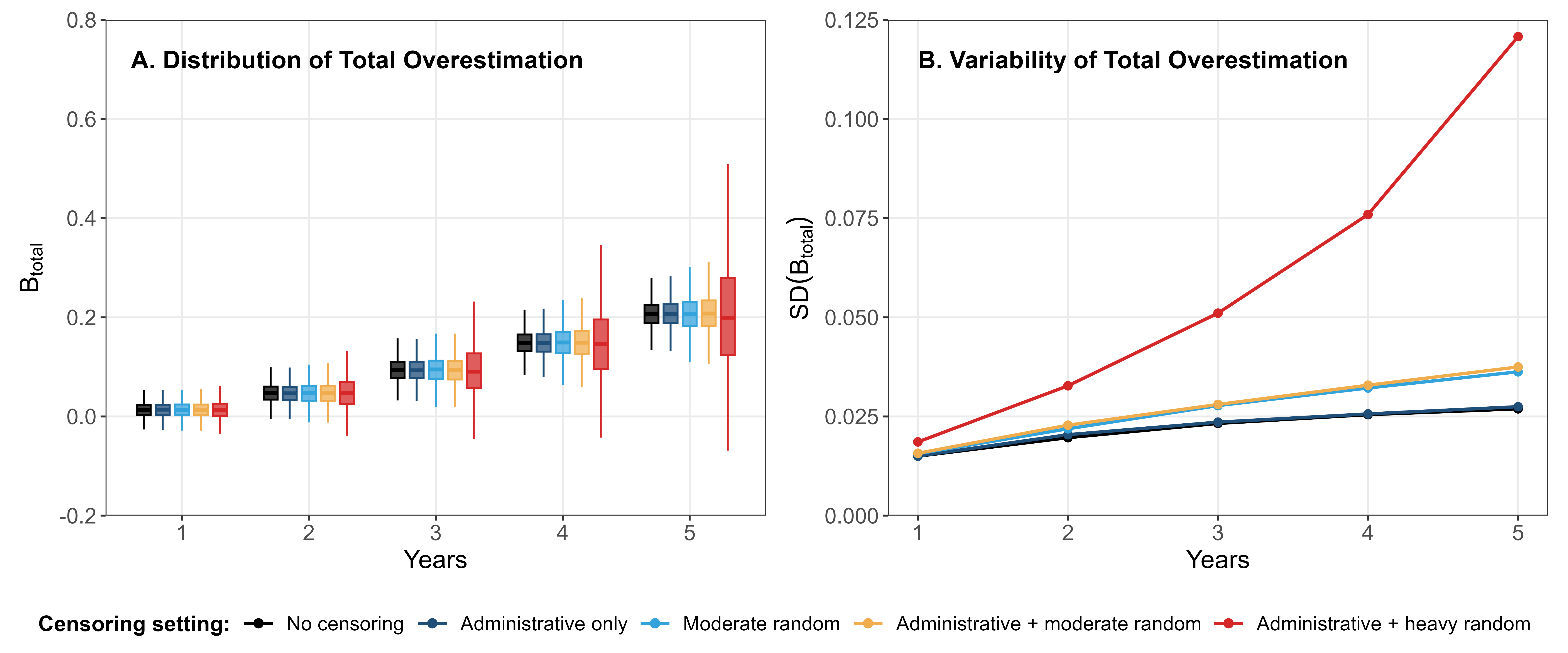} 
		\caption{Distribution of total KM overestimation (A) and corresponding variability (B) under five censoring settings. 
		Total overestimation magnitude remains stable across censoring settings, 
		whereas variability increases markedly with censoring intensity.
		These findings indicate that independent censoring primarily affects precision 
		rather than the underlying overestimation magnitude.} 
		\label{fig:censoring_boxplots} 
	\end{figure*}

	The Product Limit Error Equivalence theorem predicts that 
	independent censoring should primarily affect the precision of total overestimation estimation 
	rather than the magnitude of total KM overestimation. 
	To evaluate this prediction, we simulated five censoring settings 
	ranging from complete follow-up to heavy random censoring. 
	Figure~\ref{fig:censoring_boxplots} summarises the resulting distributions of total overestimation. 
	
	Across all censoring settings, total overestimation remained remarkably stable. 
	At five years, mean total overestimation ranged
	from approximately 0.207 in the uncensored setting to 0.209 under the heaviest configuration
	(Appendix~\ref{bmsec:censoring_stress}). 
	In contrast, estimator variability increased markedly with censoring intensity. 
	The standard deviation of total overestimation increased 
	from approximately 0.027 to 0.121, representing a 4.5-fold increase in variability. 
	
	These results demonstrate a clear separation between magnitude and precision. 
	Independent censoring has little effect on underlying overestimation, 
	but considerably reduces the precision with which that overestimation can be estimated. 
	The observed behaviour is therefore consistent with the prediction of Product Limit Error Equivalence.
	
\section{A Practical Interpretation Framework}
	 	\label{sec:practical_relevance}
		
		Two practical questions remain unresolved.
		First, given that both event ordering and competing event frequency influence KM overestimation, 
		how their effects can be quantified and their relative importance determined.
		Secondly, how competing risks relevance should be assessed in routine applications and, 
		consequently, when cumulative incidence methods should be preferred to na\"ive KM risks.
		
\subsection{Empirical Characterisation of Cause-Specific Overestimation}	
		
		To address the first question, 
		we investigated how cause-specific KM overestimation at horizon (end of follow-up) relates to 
		total overestimation (TOD) at horizon, competing event frequency, and event ordering measured by Order Index (OI)
		across the simulation design from Section~\ref{subsec:phase4_simulation}.
		To characterise this relationship,
		a sequence of increasingly complex models was fitted using the simulated datasets. 
		Simulation performance was evaluated using adjusted $R^2$ and 
		external validation using the six benchmark datasets. 
		Complete model results are reported in Appendix~\ref{app:phase4_design}.
		Table~\ref{tab:models_adjR2} summarises the principal findings. 
	
		\begin{table}[htbp]
			\centering
			\caption{
			Models relating cause-specific KM overestimation to
			Total Overestimation Diagnostic (TOD), Order Index (OI), and competing event frequency. 
			Simulation performance is measured by adjusted $R^2$; 
			external validation is evaluated using the benchmark datasets from Table~\ref{tab:benchmark_datasets}. 
			MAE denotes mean absolute prediction error and Correlation$=\operatorname{corr}(\textnormal{prediction}, \Biask(t_{\max}))$.}
			\label{tab:models_adjR2}
			\footnotesize
			\begin{tabular}{lccc}
				\toprule  & \multicolumn{1}{c}{Simulation} & \multicolumn{2}{c}{Validation} \\ 
				Linear Model predicting $\Biask$ &	Adjusted $R^2$ & Correlation & MAE \\
				\midrule
				Frequency & 0.246 & 0.695 & 0.231\\
				OI              & 0.290 & -0.150 & 0.253\\
				TOD          & 0.477 & {\bf 0.986} & 0.017\\
				\midrule
				Frequency + OI & 0.512 & 0.650 &0.237\\
				TOD + Frequency & 0.639 & 0.859 & 0.049\\
				TOD $\times$ Frequency & 0.862 & 0.815 & 0.029\\
				TOD + OI   & 0.737 & 0.815 & 0.026 \\
				TOD $\times$ OI   & 0.806 & {\bf 0.987} & {\bf 0.007} \\
				\midrule
				TOD + Frequency + OI & 0.883 & 0.933 & 0.043\\
				TOD $\times$ Frequency $\times$ OI &  0.953 & 0.941 & 0.029 \\
				\bottomrule
			\end{tabular}
		\end{table}
	
		The benchmark validation results suggest two particularly useful empirical relationships
		for characterising cause-specific overestimation:
		\begin{equation}
			\label{eq:model_TOD}
			\Biask(t_{\max}) \simeq 0.5\, TOD ,
		\end{equation}
		\begin{equation}
			\label{eq:model_TOD_OI}
			\Biask(t_{\max}) \simeq \left( 0.13 + 0.79\, OI \right)\, TOD \;. 
		\end{equation}
	
		Several more complex models achieved higher adjusted $R^2$ values on the simulated data. 
		However, the TOD$\times$OI relationship generalised most effectively to the benchmark datasets (MAE=0.007), 
		outperforming models based on competing event frequency alone, OI alone, TOD alone, and more complex model specifications. 
		These results suggest that the combined information contained in TOD and OI 
		captures the principal determinants of cause-specific overestimation, 
		whereas competing event frequency provides comparatively little additional explanatory value. 
		
		Since the TOD$\times$OI relationship achieved the strongest external validation performance, 
		it provides an empirical 
		characterisation of the principal determinants of cause-specific overestimation.
		Specifically, Equation~\eqref{eq:model_TOD_OI} can be rewritten as
		\begin{equation}
			\label{eq:model_Bias_TOD_ratio}
		\boxed{
			\frac{ \Biask(t_{\max}) }{ TOD(t_{\max}) }  \simeq 0.13 + 0.79\, OI\;, 
		}	
		\end{equation}
		which quantifies how the share of total overestimation 
		corresponding to the event of interest depends on OI.
	
		Equation~\eqref{eq:model_TOD} suggests that, in the absence of additional information, 
		cause-specific overestimation is approximately one half of total overestimation.
		Equation~\eqref{eq:model_Bias_TOD_ratio} refines this relationship through OI, 
		providing a quantitative interpretation of how event ordering influences cause-specific overestimation.
	
		This interpretation
		is closely related to the classical competing risks framework. 
		Lower OI values imply that events of interest occur earlier 
		and the risk set depletion due to competing events occurs later, 
		reducing cause-specific KM overestimation. 
		Conversely, higher OI values imply earlier risk set depletion
		so that cause-specific overestimation represents a larger share of total overestimation.
		
		Finally, note that the relationships \eqref{eq:model_TOD} and \eqref{eq:model_Bias_TOD_ratio} 
		generalised well to the benchmark datasets despite considerable variation in sample size and censoring frequency.
		This suggests that the relationships identified by the simulation study (n=500 and 20\% of censoring) 
		are not highly sensitive to the specific sample size and censoring configuration used during model development.
		
	\subsection{Implications for Assessing Competing Risks Relevance}	
	
		We now return to the practical question motivating this work: 
		how should competing risks relevance be assessed in routine applications and, consequently, 
		when should cumulative incidence methods be preferred to naive KM risks?
		
		We suggest a simple assessment workflow. 
		First, compute TOD and its confidence interval. 
		If TOD is sufficiently small,
		competing risks overestimation is unlikely to have practical consequences.
		If TOD is not negligible, 
		further analysis should depend on the scientific objective. 
		If interest lies in the competing risks system as a whole, 
		a non-negligible TOD reveals that competing risks are practically relevant.
		If interest centres on a particular event,
		cause-specific overestimation should be examined directly 
		through KM--CIF comparisons for the event of interest.
		The choice of what constitutes a negligible TOD is application dependent 
		and should reflect the maximum discrepancy investigators 
		are willing to tolerate between KM and cumulative incidence estimates.
		
		The rationale for beginning with TOD is straightforward. 
		TOD quantifies total KM overestimation across the competing risks system, 
		whereas cause-specific overestimation depends on the choice of event of interest.
		Since cause-specific overestimation cannot exceed total overestimation, 
		a small TOD implies uniformly small cause-specific overestimation for all event types. 
		Consequently, when TOD is judged negligible,
		competing risks are unlikely to materially affect any conclusions
		regardless of the endpoint under study.
		
		The role of OI is interpretive rather than diagnostic. 
		TOD and KM--CIF comparisons quantify the magnitude of overestimation, 
		at total and cause-specific levels,
		whereas OI helps explain why the latter overestimation occurs.
		By quantifying temporal event allocation, 
		OI summarises the risk set depletion mechanisms that drive cause-specific overestimation.
		
		Together, TOD and OI provide a practical framework 
		for assessing competing risks relevance.
		TOD quantifies the magnitude of total KM overestimation, 
		whereas OI explains how temporal event allocation 
		influences cause-specific overestimation for a particular event of interest. 
		The results of this study therefore suggest 
		that competing risks relevance depends on both 
		the magnitude of total KM overestimation and 
		the temporal allocation of event types, 
		rather than on competing event frequency alone.
		
\section{Discussion} \label{sec:discussion} 
	
	This paper developed a framework for understanding competing risks overestimation. 
	Rather than treating overestimation as 
	a qualitative consequence of applying Kaplan--Meier methods in the presence of competing risks, 
	the proposed framework allows overestimation to be 
	explained, quantified, subjected to formal inference, and interpreted, 
	thereby providing a practical basis for assessing the relevance of competing risks in routine analyses.
	
	The first question addressed by the framework is: What determines KM overestimation?
	Previous discussions of competing risks overestimation have largely focused 
	on its existence and on the role of competing event frequency \cite{pepe1993kaplan,pintilie2006competing,austin2017accounting}. 
	By contrast, the present work showed that cause-specific overestimation possesses 
	an explicit temporal structure arising from the interaction between competing events and events of interest. 
	The ordering results identified attainable lower and upper overestimation boundaries, 
	characterised the local accumulation mechanism, 
	introduced the Order Index, and 
	provided a complete ordering that maximises horizon cause-specific KM overestimation. 
	These results establish event ordering as an independent determinant of overestimation magnitude.
	
	The empirical analyses reinforced this interpretation.
	Benchmark datasets demonstrated considerable heterogeneity in overestimation 
	despite similar competing event frequencies, 
	while simulation experiments showed that 
	ordering effects become increasingly pronounced as competing event frequency increases. 
	These findings indicate that temporal event allocation is 
	an important source of heterogeneity in the practical behaviour of KM overestimation. 
	They also highlight an important distinction between cause-specific and total overestimation: 
	configurations producing maximal cause-specific overestimation for one event type 
	do not necessarily produce maximal total overestimation. 
	Therefore, cause-specific and total overestimation perspectives should not be regarded as interchangeable.
		
	The second question addressed by the framework is: How much overestimation exists?
	The Total Overestimation Diagnostic (TOD) converts competing risks overestimation 
	into a directly measurable quantity that can be obtained from routine survival analysis output. 
	Importantly, Structural Positivity showed that total KM overestimation is 
	an unavoidable feature of realised competing risks settings.
	TOD therefore quantifies the magnitude of this unavoidable overestimation.
	The benchmark datasets demonstrated substantial heterogeneity in TOD, 
	even among settings with similar competing event frequencies, 
	reinforcing the distinction between competing event frequency and actual overestimation. 
	By making total overestimation observable, 
	TOD provides a practical audit metric and establishes a quantitative basis 
	for assessing the relevance of competing risks in routine analyses.
	
	The third question addressed by the framework is: 
	How should uncertainty in competing risks overestimation be quantified?
	AJ Error Equivalence showed that inference for empirical cause-specific overestimation 
	inherits directly from the Aalen--Johansen estimator. 
	At the competing risks setting level, Product Limit Error Equivalence demonstrated that 
	inference for TOD reduces exactly, rather than asymptotically, to ordinary overall survival estimation. 
	Consequently, confidence intervals, standard errors, and uncertainty quantification for competing risks overestimation 
	can be obtained using existing survival analysis methodology, 
	without covariance modelling, bootstrap resampling, or inferential procedures specific to competing risks.
	This reduction in inferential complexity substantially lowers the practical barrier 
	to uncertainty quantification for competing risks overestimation.
	
	The framework concluded by addressing two practical questions:
	How can cause-specific overestimation be characterised?
	When are competing risks practically important? \\
	In the absence of additional information, 
	cause-specific overestimation is approximately one half of total overestimation on average. 
	Incorporating event ordering substantially refines this relationship. 
	TOD quantifies how much competing risks overestimation exists, 
	whereas the Order Index (OI) 
	provides an interpretation of the resulting cause-specific overestimation.\\
	Together, TOD and OI provide a practical framework for assessing competing risks relevance.
	Since cause-specific overestimation cannot exceed total overestimation, 
	small TOD values imply uniformly small cause-specific overestimation across all event types.
	Accordingly, we propose to first assess whether TOD is negligible.
	If TOD is sufficiently small,
	cause-specific overestimation must also be small for every event type, and
	competing risks are unlikely to materially affect substantive conclusions.
	If TOD is not negligible and the study interest lies on a particular event, 
	cause-specific overestimation should be quantified directly 
	through KM--CIF comparisons for that event.
	The role of OI is then interpretive rather than diagnostic.
	This part of the framework provides the practical counterpart 
	to the event ordering theory developed in Section~\ref{sec:event_ordering}.
	
	Several limitations should be acknowledged.
	The theoretical results were developed under a standard competing risks setting 
	with independent right censoring and without delayed entry, 
	interval censoring, recurrent event processes, or time dependent covariates. 
	The ordering results describe structural properties of 
	observed competing risks trajectories and should not be interpreted causally. 
	Finally, although the theoretical contributions apply to an arbitrary number of event types, 
	the empirical illustrations focused primarily on settings with two competing event types. 

	Future work may extend the framework beyond standard independent censoring. 
	Particularly important directions include the development of covariate adjusted versions of TOD, 
	extensions to multi-state and progressive disease processes, and 
	characterisation of the event allocation that maximises total overestimation across an entire competing risks setting. 
	
	The proposed framework allows  the relevance of competing risks
	to be assessed explicitly rather than merely inferred. 
	By quantifying total overestimation (TOD), 
	directly evaluating cause-specific overestimation when necessary, 
	and using temporal allocation of events (OI) to interpret the resulting discrepancies,
	the practical relevance of competing risks can be evaluated directly from observed data. 
	TOD provides the assessment; 
	KM--CIF provides event-specific quantification;
	OI provides the interpretation.
	This provides a principled basis for determining when 
	na\"ive KM risks remain reasonable approximations of real-world event probabilities and when 
	cumulative incidence methods become necessary. 
	In this sense, competing risks overestimation becomes 
	not simply a methodological limitation of KM estimation, but 
	a routinely measurable quantity that can inform analytic and reporting decisions.

\bmsubsection*{Software and Reproducibility. } 

	All empirical analyses were conducted using R version~4.4.1 \cite{Rsoftware}. 
	Survival and competing risks estimation used the 
	\texttt{survival} version~3.8.6 and \texttt{cmprsk} version~2.2.12 packages, 
	respectively \cite{survival-package, cmprsk-package}. 
	The proposed TOD and OI methods, complete analysis scripts, 
	and reproduction instructions are available from the 
	\texttt{kmaudit} project repository at \url{https://github.com/fespuny/kmaudit}. 
	The version of \texttt{kmaudit} that includes the reproduction materials 
	corresponding to this article is 
	permanently archived on Zenodo \cite{fep2026kmaudit}.

\bmsubsection*{Data Availability Statement} 

	The analysis data that support the findings of this study are
	openly available in the versioned \texttt{kmaudit} archive on Zenodo at 
	\href{https://doi.org/10.5281/zenodo.24134374} {doi:10.5281/zenodo.24134374}
	 \cite{fep2026kmaudit}. 
	The archive includes the simulation level results and derived analysis data 
	required to reproduce the reported figures, tables, model fits, and validation results. 
	Benchmark analyses use publicly available datasets which are distributed 
	through the R packages cited in this article and remain subject to 
	their original licences and terms of use.
		
\bmsubsection*{Financial Disclosure}
	
	None reported.
	
\bmsubsection*{Conflicts of Interest}
	
	The authors declare no conflicts of interest.
	
\bibliographystyle{ama} 

\bibliography{references}
	
\appendix

\counterwithin{table}{section} 
\renewcommand{\thetable}{\thesection\arabic{table}} 
\counterwithin{figure}{section} 
\renewcommand{\thefigure}{\thesection\arabic{figure}}

\bmsection{Supplementary Results for Event Ordering} 
	\label{app:sec:event_ordering}
	
	This appendix provides technical results and proofs supporting the event ordering theory developed in Section~\ref{sec:event_ordering}. 
	Throughout, observed failure times are assumed distinct, $d(t_i)\le1$, and all observed event types other than type $k$ are aggregated into the competing event label $E_c$. 
	This aggregation does not affect the arguments, which depend only on whether an observed failure is of type $k$ or a competing type.
	\begin{itemize}
	\item Appendix~\ref{app:local_inflation}
		provides the complete derivation of the Local Empirical Overestimation Inflation
		Formula in Equation~\eqref{eq:local_bias_inflation}.
	\item Appendix~\ref{app:uncensored_bound}
		discusses intervening censoring structures and derives the closed-form
		uncensored upper bound for maximal KM overestimation.
	\end{itemize}

\bmsubsection{Derivation of the Local Empirical Overestimation Inflation Formula} 
	\label{app:local_inflation} 

	This appendix derives the exact local increase in empirical cause-specific KM overestimation 
	created by transposing an adjacent pair of failure events: $(E_k,E_c)$ into $(E_c,E_k)$.  
	Consider two consecutive failure times $t_a<t_b$ with no intervening failures 
	and with $c\ge0$ censored observations occurring between them. 
	Let \[ r_a=r(t_a^-), \qquad r_b=r(t_b^-)=r_a-1-c. \] 
	Define \[ D_{\mathrm{prev}} = 1-\hat S_{\mathrm{KM},k}(t_a^-) - \hat F_k^{AJ}(t_a^-) \] 
	and \[ \hat S_{\mathrm{PL,prev}} = \hat S_{\mathrm{PL}}(t_a^-). \] 
	
\bmsubsubsection{Original Allocation: \texorpdfstring{$(E_k,E_c)$}{(Ek,Ec)}} 

	Under the original sequence $(E_k,E_c)$, the event of type $k$ occurs at $t_a$. 
	The KM complement therefore updates as 
	\begin{align} 
	1-\hat S_{\mathrm{KM},k}(t_b^+) &= 1-\hat S_{\mathrm{KM},k}(t_a^-) \left( 1-\frac1{r_a} \right) \\ 
	                                                      &= 1-\hat S_{\mathrm{KM},k}(t_a^-) + \frac{\hat S_{\mathrm{KM},k}(t_a^-)}{r_a}. 
	\end{align} 
	Similarly, the AJ estimator updates as 
	\begin{equation} 
		\hat F_k^{AJ}(t_b^+) = \hat F_k^{AJ}(t_a^-) + \frac{\hat S_{\mathrm{PL,prev}}}{r_a}. 
	\end{equation} 
	Consequently, 
	\begin{equation} 
		\hat B_k(t_b^+) = D_{\mathrm{prev}} + \frac{ \hat S_{\mathrm{KM},k}(t_a^-) - \hat S_{\mathrm{PL,prev}} }{r_a}.
	\label{eq:bias_original_app} 
	\end{equation} 
	
	\bmsubsubsection{Transposed Allocation: \texorpdfstring{$(E_c,E_k)$}{(Ec,Ek)}} 
	Under the transposed sequence $(E_c,E_k)$, the competing event occurs first. 
	The KM complement therefore becomes 
	\begin{align} 
		1-\hat S'_{\mathrm{KM},k}(t_b^+) &= 1-\hat S_{\mathrm{KM},k}(t_a^-) \left( 1-\frac1{r_b} \right) \\ 
		&= 1-\hat S_{\mathrm{KM},k}(t_a^-) + \frac{\hat S_{\mathrm{KM},k}(t_a^-)}{r_b}. 
	\end{align} 
	For the AJ estimator, the competing event first reduces overall product limit survival to 
	\[ \hat S_{\mathrm{PL,prev}} \left( 1-\frac1{r_a} \right). \] 
	The subsequent cause-$k$ event then contributes 
	\[ \frac{ \hat S_{\mathrm{PL,prev}} (r_a-1) }{ r_a r_b }. \] 
	Hence 
	\begin{equation} 
		\hat F_k^{AJ\,\prime}(t_b^+) = \hat F_k^{AJ}(t_a^-) + \frac{ \hat S_{\mathrm{PL,prev}} (r_a-1) }{ r_a r_b }. 
	\end{equation} 
	
\bmsubsubsection{Difference in Empirical Overestimation (Exact Local Inflation Formula)} 

	The change in empirical KM overestimation induced by the transposition is 
	\[ \Delta \hat B_k(t_b^+) = \hat B_k'(t_b^+) - \hat B_k(t_b^+). \] 
	Substituting the preceding expressions and simplifying yields 
	\[ \Delta \hat B_k(t_b^+) = \frac{ D_{\mathrm{prev}}(r_a-r_b) + \hat S_{\mathrm{PL,prev}} }{ r_a r_b}. \] 
	Since \[ r_a-r_b = 1+c, \] we obtain 
	\begin{equation} 
		\boxed{ \Delta \hat B_k(t_b^+) = \frac{ D_{\mathrm{prev}}(1+c) + \hat S_{\mathrm{PL,prev}} }{ r_a r_b} } \label{eq:local_bias_inflation_appendix} 
	\end{equation} 
	which is exactly the expression reported in 
	Equation~\eqref{eq:local_bias_inflation}.
	Because \[ D_{\mathrm{prev}}\ge0, \qquad \hat S_{\mathrm{PL,prev}}>0, \qquad r_a r_b>0, \] 
	it follows immediately that \[ \Delta \hat B_k(t_b^+) >0. \] 
	Therefore removing an inversion by an adjacent transposition
	strictly increases empirical cause-specific KM overestimation. \qed
	
\bmsubsection{Extensions and Sharp Upper Bound}\label{app:uncensored_bound}
	
\paragraph*{Intervening Censoring Structures.} 
	The event ordering results developed in Section~\ref{sec:event_ordering} 
	extend directly to cohorts subject to intervening right censoring. 
	Let $c \ge 0$ denote the number of individuals censored between two adjacent failure events occurring at times $t_a < t_b$. 
	The Local Empirical Overestimation Inflation Formula in Equation~\eqref{eq:local_bias_inflation} was derived 
	allowing arbitrary values of $c$ through the relationship $r_b=r_a-1-c$, 
	and therefore already incorporates censoring occurring between adjacent failure events. 
	Because Equation~\eqref{eq:local_bias_inflation} remains strictly positive 
	for all admissible values of $c$, $\Delta \hat B_k(t_b^+) > 0$, 
	removing an inversion by an adjacent transposition
	continues to increase 
	empirical KM overestimation in the presence of right censoring. 
	Consequently, the monotonic inversion reduction argument underlying Theorem~\ref{thm:km_risk_max} 
	remains valid under arbitrary independent right censoring structures.
	
\paragraph*{Closed-Form Evaluation of the Maximal Uncensored Overestimation.} 
	Theorem~\ref{thm:km_risk_max} identifies the event ordering that maximizes cause-specific KM risk
	at the study horizon
	and, by Corollary~\ref{cor:theoretical_bias_max}, the corresponding theoretical horizon overestimation. 
	Consider an uncensored cohort of size $N$ containing $n_k$ events of type $k$ and $n_c$ competing events. 
	Because the Aalen--Johansen estimator reduces to the observed event proportion under complete follow-up, 
	\[ \widehat F_k^{\mathrm{AJ}}(t_{\max}) = \frac{n_k}{N}. \] 
	Under the maximal-overestimation configuration 
	$\mathcal S_F^\ast = (E_c,\ldots,E_c,E_k,\ldots,E_k)$, 
	all competing events occur before all events of type $k$. 
	The corresponding KM survival estimator therefore satisfies 
	\[ \widehat S_{\mathrm{KM},k}^{\max}(t_{\max}) = \prod_{j=0}^{n_k-1} \left( 1-\frac{1}{N-n_c-j} \right) = \frac{N-n_c-n_k}{N-n_c}. \] 
	Taking complements gives 
	\[ 1-\widehat S_{\mathrm{KM},k}^{\max}(t_{\max}) = \frac{n_k}{N-n_c}. \] 
	Hence the maximal attainable empirical overestimation is 
	\begin{equation} 
		\widehat B_k^{\max}(t_{\max}) = \frac{n_k}{N-n_c} - \frac{n_k}{N} = \frac{n_k n_c}{N(N-n_c)}. \label{eq:uncensored_max_bias_app} 
	\end{equation} 
	Equation~\eqref{eq:uncensored_max_bias_app} provides a sharp closed-form upper bound 
	for empirical KM overestimation under complete follow-up. 
	The bound depends only on the numbers of events of interest and competing events, 
	and is attained precisely under the zero inversion ordering characterized by Theorem~\ref{thm:km_risk_max}.
	
\bmsection{Simulation Studies} \label{app:simulation}
	
	This appendix mainly provides supplementary details for the simulation studies reported in Section~\ref{sec:empirical_evaluation}. 
	\begin{itemize}
		\item Appendix~\ref{app:phase4_design} describes the frequency-ordering simulation design and benchmarking;
		\item Appendix~\ref{app:phase4_calibration} reports calibration and quality control results;
		\item Appendix~\ref{bmsec:censoring_stress} provides supplementary results for the censoring stress test; 
		\item Appendix~\ref{app:pl_validation} provides numerical validation of the Product Limit Error Equivalence.
	\end{itemize}
	
	\bmsubsection{Frequency \& OI Space Simulation Design and Benchmarking} \label{app:phase4_design} 
	
	The simulation experiments reported in Sections~\ref{subsec:phase4_simulation} and~\ref{sec:practical_relevance} 
	were generated using a Frequency-OI space simulation design rather than a conventional parameter space factorial design. 
	The design was developed in three stages. 
	
	First, a calibration phase was performed using a large collection of latent competing risks processes. 
	For each combination of latent Weibull parameters, 
	the generated competing event frequency and Order Index (OI) were recorded. 
	This produced an empirical calibration cloud linking latent parameter configurations to simulated observable quantities. 
	
	Second, the frequency \& ordering space generated by the calibration cloud was evaluated. 
	Preliminary experiments showed that direct parameter space sampling produced 
	highly uneven coverage of the frequency \& OI space, 
	with some combinations of competing event frequency and event ordering markedly overrepresented and others sparsely represented. 
	
	Third, an occupancy aware sampling procedure was developed to improve coverage of the simulated space. 
	Calibration points were sampled from the empirical calibration cloud using inverse occupancy weighting, 
	thereby increasing the representation of sparsely populated regions while preserving the diversity of calibration configurations. 
	The weighting scheme was tuned using the occupancy diagnostics reported in Appendix~\ref{app:phase4_calibration}. 
	
	Using this approach, simulated datasets were generated with target sample size $N=500$ and approximately 20\% independent right censoring. 
	A total of $N_{total}=100\times 800=80,000$ datasets were produced. 
	The resulting simulation design achieved broad coverage of the frequency \& ordering space 
	and allowed empirical investigation of KM overestimation beyond the limited set of ordering settings considered in earlier experiments. 
	For each simulated dataset, the following quantities were recorded: 
	\begin{itemize} 
		\item censoring proportion, 
		\item competing event frequency, 
		\item Order Index, 
		\item KM risks, 
		\item cumulative incidence functions, 
		\item empirical cause-specific KM overestimation, 
		\item empirical total KM overestimation (TOD). 
	\end{itemize} 
	
\paragraph*{Model benchmarking and sensitivity analysis. }
	The resulting datasets were used to evaluate how frequency, ordering, TOD, and censoring 
	jointly influence competing risks overestimation (Table~\ref{tab:models_adjR2}).
	Table~\ref{tab:all_models} reports the complete model search considered 
	during development of the practical interpretation framework presented in Section~\ref{sec:practical_relevance}.
	Although the three-factor interaction model achieved the highest simulation adjusted $R^2$, 
	the TOD$\times$OI model yielded the strongest benchmark validation performance (MAE=0.007) 
	and therefore formed the basis of the developed interpretation framework.

	As a sensitivity analysis, all models were re-fitted using occupancy balanced undersampling, 
	with 250 simulated datasets drawn for each Frequency by OI cell in a 10 $\times$ 10 discretisation grid. 
	The resulting model coefficient, adjusted $R^2$ and benchmark validation values were nearly identical 
	to those obtained under the original design. 
	In particular, the TOD and TOD $\times$ OI models retained both their coefficient structure and relative performance, 
	suggesting that the proposed interpretation framework is not sensitive to the occupancy structure of the simulated space.

	\begin{table*}[htbp] 
		\centering 
		\caption{Full model comparison, leading to the TOD \& OI interpretation framework. 
			All models predict cause-specific overestimation at horizon: $\Biask(t_{\max})$.
			Model Coefficients columns indicate the regression terms included in each model. 
			Simulation performance is measured by adjusted $R^2$,
			External validation is evaluated using the six benchmark datasets from Table~\ref{tab:benchmark_datasets}. 
			MAE denotes mean absolute prediction error;
			Corr.~= $\operatorname{corr}\left(\textnormal{prediction}, \Biask(t_{\max}) \right)$; 
			OI: Order Index; TOD: Total Overestimation Diagnostic at $t=t_{\max}$.}
		\label{tab:all_models} 
		\footnotesize 
		\resizebox{\textwidth}{!}{%
		\begin{tabular}{lcccccccc|ccc} 
			\toprule & \multicolumn{8}{c}{Model Coefficients} & \multicolumn{1}{c}{Simulation} & \multicolumn{2}{c}{Validation} \\ 
			Model & Constant & Freq & OI & TOD & Freq$\times$OI & TOD$\times$Freq & TOD$\times$OI & TOD$\times$Freq$\times$OI & Adj. $R^2$ & Corr. & MAE \\ 
			\midrule
			Frequency & 0.105 & 0.404 & & & & & & & 0.246 & 0.695 & 0.231 \\ 
			OI & 0.086 & & 0.446 & & & & & & 0.290 & -0.150 & 0.253 \\ 
			TOD & -0.025 & & & 0.560 & & & & & 0.477 & 0.986 & 0.017 \\ 
			\midrule 
			Frequency + OI & -0.096 & 0.384 & 0.428 & & & & & & 0.512 & 0.650 & 0.237 \\ 
			Frequency $\times$ OI & 0.200 & -0.213 & -0.182 & & 1.211 & & & & 0.676 & 0.750 & 0.226 \\ 
			TOD + Frequency & -0.163 & 0.331 & & 0.513 & & & & & 0.639 & 0.859 & 0.049 \\ 
			TOD $\times$ Frequency & 0.102 & -0.213 & & -0.099 & & 1.203 & & & 0.862 & 0.815 & 0.029 \\ 
			TOD + OI & -0.224 & & 0.423 & 0.543 & & & & & 0.737 & 0.815 & 0.026 \\ 
			TOD $\times$ OI & -0.001 & & -0.014 & 0.134 & & & 0.794 & & 0.806 & \textbf{0.987} & \textbf{0.007} \\ 
			\midrule 
			TOD + Frequency + OI & -0.348 & 0.313 & 0.410 & 0.499 & & & & & 0.883 & 0.933 & 0.043 \\ 
			TOD + Frequency $\times$ OI & -0.251 & 0.171 & 0.260 & 0.454 & 0.301 & & & & 0.890 & 0.918 & 0.053 \\ 
			TOD $\times$ Frequency + OI & -0.123 & -0.028 & 0.286 & 0.112 & & 0.767 & & & 0.951 & 0.926 & 0.025 \\ 
			TOD $\times$ OI + Frequency & -0.274 & 0.281 & 0.292 & 0.392 & & & 0.217 & & 0.886 & 0.938 & 0.037 \\ 
			TOD $\times$ Frequency $\times$ OI & -0.055 & -0.116 & 0.167 & 0.046 & 0.194 & 0.807 & 0.117 & -0.129 & 0.953 & 0.941 & 0.029 \\
			\bottomrule 
		\end{tabular} 
		}
	\end{table*}

	To examine practical predictive performance in detail, 
	the strongest benchmark performing models were applied directly to the six benchmark datasets. 
	Table~\ref{tab:relevance_validation} compares observed and predicted cause-specific overestimation.
	The principal validation results are summarised in the main text Table~\ref{tab:models_adjR2}.
	
	\begin{table*}[htbp] \centering 
		\caption{ External validation of competing risk relevance models using the benchmark datasets from Table~\ref{tab:benchmark_datasets}.
			Negative predictions were set to zero given that overestimation is non-negative.
			Absolute errors are shown relative to the observed cause-specific KM overestimation. } 
		\label{tab:relevance_validation} 
		\footnotesize 
		\begin{tabular}{lccccccc} 
		\toprule & \multicolumn{1}{c}{Observed} & \multicolumn{2}{c}{TOD} & \multicolumn{2}{c}{TOD $\times$ OI} & \multicolumn{2}{c}{TOD $\times$ Frequency $\times$ OI} \\ 
		Dataset & $\Biask(t_{\max})$ & Pred. & Error & Pred. & Error & Pred. & Error \\ 
		\midrule 
		PBC & 0.031 & 0.012 & 0.019 & 0.023 & 0.007 & 0.018 & 0.013 \\ 
		MELANOMA & 0.016 & 0.002 & 0.014 & 0.019 & 0.003 & 0.044 & 0.028 \\ 
		UDCA & 0.004 & 0.000 & 0.004 & 0.000 & 0.004 & 0.047 & 0.043 \\ 
		MYELOID & 0.077 & 0.051 & 0.026 & 0.075 & 0.003 & 0.102 & 0.025 \\ 
		MGUS & 0.146 & 0.135 & 0.010 & 0.145 & 0.001 & 0.198 & 0.052 \\ 
		TRANSPLANT & 0.081 & 0.054 & 0.026 & 0.059 & 0.022 & 0.092 & 0.011 \\ 
		\midrule 
		Mean Absolute Error &  & & 0.017 & & \textbf{0.007} & & 0.029 \\
		\bottomrule 
		\end{tabular} 
	\end{table*}

\bmsubsection{Calibration and Coverage Assessment}\label{app:phase4_calibration} 
	
	The simulation design was evaluated using calibration and occupancy diagnostics. 
	Calibration quality was assessed by comparing simulation generated quantities with their corresponding calibration targets. 
	
	Across the complete simulation design, the mean absolute frequency calibration error was approximately 0.016, 
	while the mean absolute Order Index calibration error was approximately 0.025. 
	These values indicate that the calibration procedure successfully generated datasets 
	spanning the intended regions of the frequency--ordering space. 
	
	Coverage of the calibration target space was assessed using a $20\times20$ occupancy grid 
	defined by observed competing event frequency and observed Order Index. 
	Every occupancy cell contained simulated datasets, indicating complete space coverage without occupancy voids (at 0.05 resolution). 
	The median occupancy exceeded 3,800 datasets per cell and the occupancy coefficient of variation was approximately 0.15, 
	representing a clear improvement over earlier simulation designs. 
	To avoid excessive concentration of simulations within a small subset of calibration points, 
	several diagnostics of calibration cloud diversity were examined. 
	These included the effective sample size, the distribution of calibration point usage, 
	and the coefficient of variation of calibration point frequencies. 
	The adopted occupancy aware weighting scheme maintained high calibration diversity while improving realised space coverage. 
	
	The simulation design achieved three objectives simultaneously: 
	\begin{enumerate} 
		\item accurate calibration of competing event frequency; 
		\item accurate calibration of event ordering; 
		\item broad and approximately uniform coverage of the joint frequency \& ordering space. 
	\end{enumerate} 
	Consequently, the simulation study provides a
	broad and balanced coverage of the frequency-ordering space.
		
\bmsubsection{Additional Results for the Censoring Stress Test} \label{bmsec:censoring_stress}
	
	This section provides detailed numerical summaries for 
	the censoring settings investigated in Section~\ref{subsec:censoring_stress}. 
	Table~\ref{tab:censoring_stress} contains both measures of total overestimation 
	and the diagnostic quantities used in the numerical validation 
	reported subsequently in Appendix~\ref{app:pl_validation}. 
	Observed censoring frequencies closely matched their intended targets 
	throughout the simulation study. 
	Detailed calibration summaries for resulting censoring levels and mean follow-up times 
	are available from the simulation quality control outputs.
	Increasing censoring substantially reduced observed follow-up time and increased variability, 
	while having little effect on the underlying total overestimation signal.
	
	\begin{table*}[htbp] \centering 
	\caption{ 
		Monte Carlo summary statistics based on 2,000 simulation replications. 
		Mean $B_{\mathrm{total}}$ and SD $B_{\mathrm{total}}$ denote the Monte Carlo mean and standard deviation of total KM overestimation. 
		SD Dia.~Error denotes the standard deviation of $\widehat B_{\mathrm{total}}-B_{\mathrm{total}}$, 
		SD PL Error denotes the standard deviation of $\widehat S_{\mathrm{PL}}-S$, 
		Max SD Difference denotes $\left|SD(\widehat B_{\mathrm{total}}-B_{\mathrm{total}}) -SD(\widehat S_{\mathrm{PL}}-S)\right|$, 
		Corr. denotes the empirical correlation between $\widehat B_{\mathrm{total}}-B_{\mathrm{total}}$ and $\widehat S_{\mathrm{PL}}-S$, and 
		Max Identity Error denotes the maximum absolute theorem residual \( \Delta = (\widehat B_{\mathrm{total}}-B_{\mathrm{total}}) - (\widehat S_{\mathrm{PL}}-S). \) 
	} \label{tab:censoring_stress} 
	\resizebox{\textwidth}{!}{%
	\begin{tabular}{lcccccccc} 
		 \toprule Setting & Time & Mean $B_{\mathrm{total}}$ & SD $B_{\mathrm{total}}$ & SD Dia.~Error & SD PL Error & Max SD Diff. & Corr. & Max Identity Error \\
		 \multirow{5}{*}{No censoring} 
		 & 1 & 0.0134 & 0.0150 & 0.0133 & 0.0133 & 0 & 1.0000 & 0 \\
		 & 2 & 0.0471 & 0.0197 & 0.0154 & 0.0154 & 0 & 1.0000 & 0 \\ 
		 & 3 & 0.0941 & 0.0232 & 0.0161 & 0.0161 & $1.39\times10^{-17}$ & 1.0000 & $8.33\times10^{-17}$ \\ 
		 & 4 & 0.1487 & 0.0255 & 0.0156 & 0.0156 & $3.12\times10^{-17}$ & 1.0000 & $5.55\times10^{-17}$ \\ 
		 & 5 & 0.2070 & 0.0269 & 0.0145 & 0.0145 & $5.20\times10^{-18}$ & 1.0000 & $1.11\times10^{-16}$ \\ 
		 \midrule 
		 \multirow{5}{*}{Administrative only} 
		 & 1 & 0.0136 & 0.0150 & 0.0133 & 0.0133 & 0 & 1.0000 & 0 \\ 
		 & 2 & 0.0470 & 0.0204 & 0.0160 & 0.0160 & 0 & 1.0000 & 0 \\ 
		 & 3 & 0.0937 & 0.0236 & 0.0163 & 0.0163 & $1.39\times10^{-17}$ & 1.0000 & $8.33\times10^{-17}$ \\ 
		 & 4 & 0.1487 & 0.0256 & 0.0156 & 0.0156 & $3.12\times10^{-17}$ & 1.0000 & $5.55\times10^{-17}$ \\ 
		 & 5 & 0.2075 & 0.0275 & 0.0148 & 0.0148 & $3.47\times10^{-18}$ & 1.0000 & $1.11\times10^{-16}$ \\ 
		 \midrule 
		 \multirow{5}{*}{Moderate random} 
		 & 1 & 0.0134 & 0.0156 & 0.0138 & 0.0138 & 0 & 1.0000 & 0 \\ 
		 & 2 & 0.0469 & 0.0219 & 0.0172 & 0.0172 & 0 & 1.0000 & 0 \\ 
		 & 3 & 0.0943 & 0.0277 & 0.0193 & 0.0193 & 0 & 1.0000 & $8.33\times10^{-17}$ \\ 
		 & 4 & 0.1491 & 0.0322 & 0.0197 & 0.0197 & 0 & 1.0000 & $1.11\times10^{-16}$ \\ 
		 & 5 & 0.2073 & 0.0363 & 0.0197 & 0.0197 & 0 & 1.0000 & $1.67\times10^{-16}$ \\ 
		 \midrule 
		 \multirow{5}{*}{Admin.~+ moderate random} 
		 & 1 & 0.0134 & 0.0157 & 0.0139 & 0.0139 & 0 & 1.0000 & 0 \\ 
		 & 2 & 0.0472 & 0.0228 & 0.0179 & 0.0179 & 0 & 1.0000 & 0 \\ 
		 & 3 & 0.0936 & 0.0280 & 0.0194 & 0.0194 & 0 & 1.0000 & $8.33\times10^{-17}$ \\ 
		 & 4 & 0.1493 & 0.0329 & 0.0201 & 0.0201 & $3.47\times10^{-18}$ & 1.0000 & $1.11\times10^{-16}$ \\ 
		 & 5 & 0.2085 & 0.0375 & 0.0201 & 0.0201 & $3.47\times10^{-18}$ & 1.0000 & $1.67\times10^{-16}$ \\ 
		 \midrule 
		 \multirow{5}{*}{Admin.~+ heavy random} 
		 & 1 & 0.0135 & 0.0186 & 0.0165 & 0.0165 & 0 & 1.0000 & 0 \\ 
		 & 2 & 0.0476 & 0.0327 & 0.0257 & 0.0257 & 0 & 1.0000 & 0 \\ 
		 & 3 & 0.0927 & 0.0511 & 0.0354 & 0.0354 & 0 & 1.0000 & $8.33\times10^{-17}$ \\ 
		 & 4 & 0.1493 & 0.0759 & 0.0463 & 0.0463 & 0 & 1.0000 & $1.11\times10^{-16}$ \\ 
		 & 5 & 0.2086 & 0.1208 & 0.0634 & 0.0634 & 0 & 1.0000 & $2.22\times10^{-16}$ \\
		 \bottomrule 
	 \end{tabular} 
	 }
	 \end{table*}

	Table~\ref{tab:censoring_stress} provides detailed Monte Carlo summaries
	for the five censoring settings investigated in Section~\ref{subsec:censoring_stress}. 
	The most important observation is the stability of mean total overestimation across censoring settings. 
	At five years, mean total overestimation ranged only 
	from approximately 0.207 in the uncensored setting 
	to approximately 0.209 under the heaviest censoring setting. 
	In contrast, variability increased markedly as censoring intensified. 
	The standard deviation of total overestimation at five years increased 
	from approximately 0.027 in the uncensored setting 
	to approximately 0.121 under heavy censoring, 
	corresponding to an approximately 4.5-fold increase in variability.

\bmsubsection{Finite Sample Validation of Product Limit Error Equivalence} \label{app:pl_validation}
	
	The Product Limit Error Equivalence theorem (Theorem~\ref{thm:global_error_equivalence}) is an exact algebraic identity. 
	The purpose of the simulations reported here is therefore not proof, 
	but numerical confirmation of the finite sample behaviour implied by the theorem.
	
\begin{figure}[htbp] \centering \includegraphics[width=\linewidth]{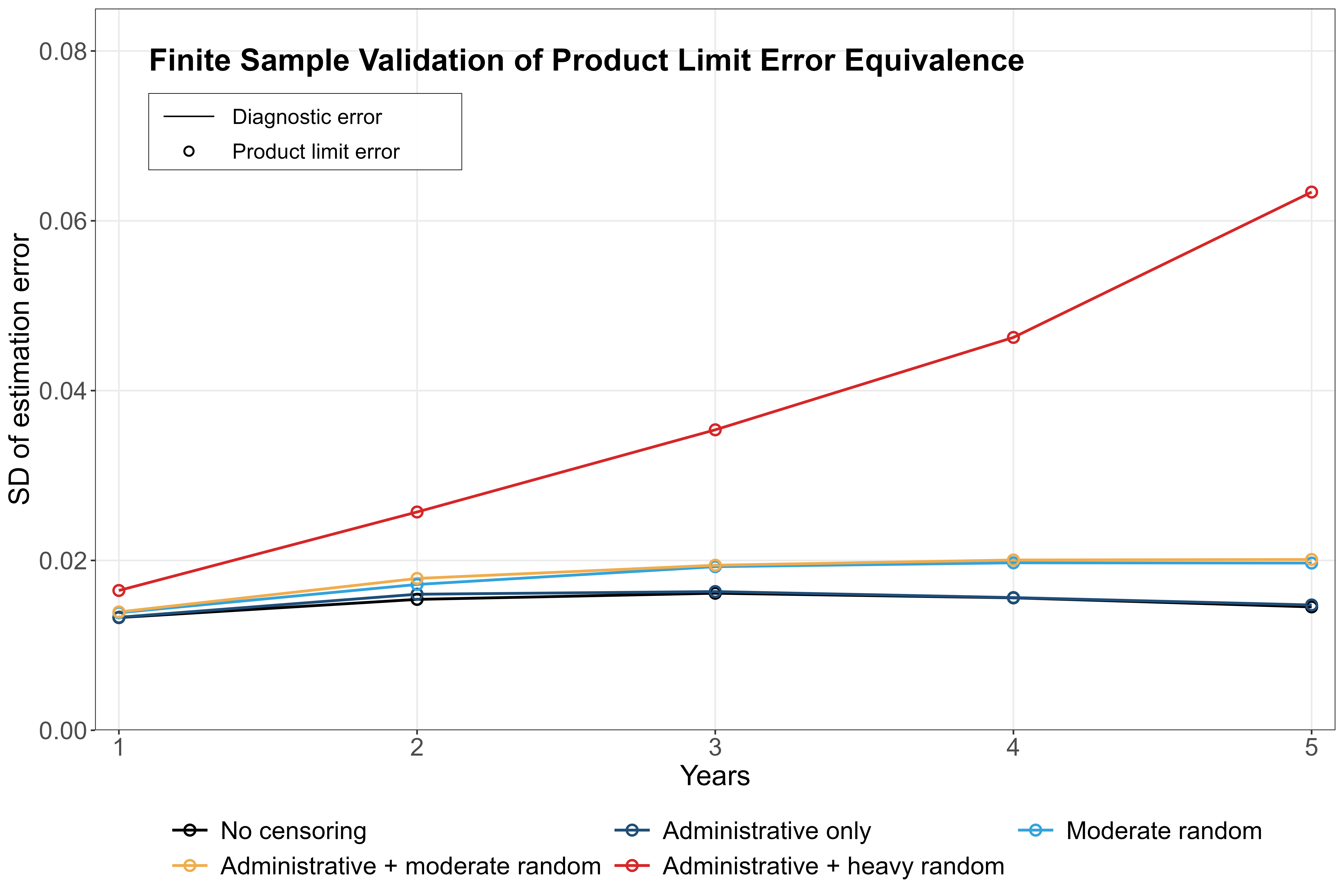} 
	\caption{Finite sample validation of the Product Limit Error Equivalence theorem. 
		Solid curves denote the standard deviation of total overestimation estimation error and 
		open circles denote the corresponding product limit survival estimation error. 
		Across all censoring settings, the two error processes are numerically indistinguishable, 
		demonstrating that inference for total overestimation behaves in finite samples 
		exactly as ordinary product limit survival inference.} 
		\label{fig:theorem4_validation} 
	\end{figure}

	We evaluated the following prediction implied by the Product Limit Error Equivalence theorem:
	that the entire finite sample error structure of total overestimation estimation should mirror 
	that of ordinary product limit survival estimation.
	Figure~\ref{fig:theorem4_validation} compares the standard deviation of the total overestimation estimation error, 
		\[ \widehat B_{\mathrm{total}}(t)-B_{\mathrm{total}}(t), \] 
	with the standard deviation of the corresponding product limit survival estimation error, 
		\[ \widehat S_{\mathrm{PL}}(t)-S(t). \] 
	
	Across all censoring settings and evaluation times, 
	the standard deviations of the two error processes 
	were numerically indistinguishable, 
	with the maximum observed difference equal to \(3.12\times10^{-17}\). 
	Furthermore, the identity residual 
	\[ \Delta(t) = \Big[ \widehat B_{\mathrm{total}}(t) - B_{\mathrm{total}}(t) \Big] - \Big[ \widehat S_{\mathrm{PL}}(t) - S(t) \Big] \] 
	remained at machine precision throughout the study. 
	The maximum absolute residual observed across all simulation runs was \(2.22\times10^{-16}\), 
	while the empirical correlation between the two error processes remained above \(0.9999999999999998\).
	 
	 Figures~\ref{fig:censoring_boxplots} and~\ref{fig:theorem4_validation} demonstrate 
	 that censoring primarily reduces precision rather than altering the magnitude of competing risks overestimation,
	 and that inference for total overestimation behaves in finite samples exactly as predicted by the Product Limit Error Equivalence theorem.

\end{document}